\documentclass[aps,prb,10pt,notitlepage,twocolumn,nofootinbib]{revtex4-1}
\usepackage{amsmath}
\usepackage{amssymb}
\usepackage{amsthm}
\usepackage{amsfonts}
\usepackage{bbm}
\usepackage{tensor}
\usepackage{dsfont}
\usepackage{graphicx}
\usepackage{subcaption}
\usepackage[section]{placeins}
\usepackage{mathrsfs}
\usepackage{stmaryrd}
\usepackage[all]{xy}
\usepackage[mathcal]{eucal}
\usepackage{verbatim}  
\usepackage{wasysym}
\usepackage{url}
\usepackage{csquotes}
\MakeOuterQuote{"}
\usepackage{ulem}
\usepackage{enumerate}
\graphicspath{{plots/}}
\newcommand{\sF}{\mathscr{F}}
\newcommand{\sG}{\mathscr{G}}
\newcommand{\sN}{\mathscr{N}}
\newcommand{\Tr}{\operatorname {Tr}}
\newcommand{\av}[1]{\left\langle #1 \right\rangle}
\newcommand{\bbI}{\mathbb{I}}
\newcommand{\dd}{{\rm d}}
\providecommand{\abs}[1]{\left\lvert #1 \right\rvert}
\providecommand{\norm}[1]{\left\lVert #1 \right\rVert}
\newcommand{\sA}{\mathscr{A}}
\newcommand{\sC}{\mathscr{C}}
\providecommand{\nCr}[2]{\left(\begin{array}{c} #1\\#2 \end{array}\right)}
\newcommand{\bbR}{\mathbb{R}}
\newcommand{\ket}[1]{\left| #1 \right\rangle}
\newcommand{\bra}[1]{\left\langle #1 \right|}
\newcommand{\inoprod}[3]{\left\langle #1\middle| #2 \middle| #3\right\rangle}
\providecommand{\hess}{\operatorname {Hess}}
\newcommand{\bbZ}{\mathbb{Z}}

\begin{document}
\title{The Sachdev-Ye-Kitaev model and free Majorana variational states}
\author{Daniel Ish}
\affiliation{RAND Corporation, Arlington, VA, 22202}
\affiliation{Physics Department, University of California, Santa Barbara, California 93106, USA}
\author{Mark Srednicki}
\affiliation{Physics Department, University of California, Santa Barbara, California 93106, USA}
\begin{abstract}
Through a mixture of analytic and numerical techniques, we explore the optimal approximation by a free Majorana state to individual disorder realizations of the Sachdev-Ye-Kitaev model, along with a generalization of it. We elucidate the properties of the known time-reversal symmetry breaking phase in the generalized model, finding strong evidence of "spin glass" order. For the Sachdev-Ye-Kitaev model itself, our results are inconclusive but suggest a similar order may be present at zero temperature.
\end{abstract}
\maketitle

\section{Introduction}
\label{intro}
The Sachdev-Ye-Kitaev model, a model of $N$ Majorana fermions subject only to quenched disordered interactions, has been an object of considerable study recently, due in part to its rich phenomenology.\cite{SY,kit_talk,subir1,MS,PhysRevD.95.046004,witten2016syklike,kitaev2018soft,polchinski,bi_xu,GPS,gross2017generalization} At finite temperature, the model is amenable to both diagrammatic and replica approaches, with the large $N$ limit reducing to a tractable integral equation which exposes an emergent (approximate) conformal symmetry. This combination of properties sparked considerable interest in the model as a means to study its holographic dual.\cite{subir1,MS,MS2,subir2,jensen}

Beyond questions of holography, this model also exhibits many properties deserving of study in their own right. To name a few, the disorder-averaged fermion two-point function is gapless and has no poles, prompting an identification of the average model as a strange metal without quasiparticles. Instabilities of this metal have also been studied as part of the general program to study exotic phases of quantum matter.\cite{bi_xu} As temperature $T\rightarrow 0$, the entropy density approaches a constant. Finally, despite the Hamiltonians of individual disorder realizations being composed purely of interactions, the replica-saddle point approach exposes that, at leading order in $N$, the disorder averaged fermion correlation functions obey Wick's theorem indicating that the average model is in some sense free at leading order.

Building on this interest, we attempt in the present work to extend considerations beyond the average model to questions about the full SYK ensemble through some modest studies of questions involving distributions and higher disorder moments of various observables. We also attempt to probe the $T=0$ physics of this model and further elaborate on studies of its instabilities. Indeed, the $T\rightarrow 0$ entropy density seen around the replica diagonal saddle point is itself evidence that the $N\rightarrow \infty$ and $T\rightarrow 0$ limits cannot be interchanged, since the entropy at $T=0$ is exactly $0$ for any finite $N$. (Technically speaking, the model could also have an extensive ground state degeneracy, but there is no evidence for this in existing exact digitalization studies.\cite{MS}) This indicates a non-analyticity in the $N\rightarrow \infty$ limit of the free energy, i.e. a phase transition.

Motivated by the observation that the average fermion correlators obey Wick's theorem and our desire to ask questions about the full distribution, we approach the problem through the variational principle applied to each individual disorder realization, taking as our mean field Hamiltonian a generic free Majorana Hamiltonian. We will see some evidence that the $T=0$ phase of the SYK model may also break time reversal symmetry, like the ordered phase observed by Bi et al. Finally, we will investigate the disorder statistics of these ordered phases, finding evidence of glassy behavior. While we advance an argument that the ordered phase observed by Bi et al is the same phase as Gaussian random Majorana fermions, we find some evidence that the $T=0$ phase of the SYK model is distinct in its disorder statistics and unlikely to be well described by our variational states.

\section{Formalism and Analytic Results}
\subsection{Setup}\label{Setup}
We take as our object of study the generalization of the Sachdev-Ye-Kitaev model studied by Bi et al\cite{bi_xu}, a model of $N$ randomly interacting Majorana fermions $\eta_i$ with Hamiltonian
\begin{align}
H &= H_0 + u H_u\\
H_0 &= \frac{1}{4!}\sum_{ijkl}^NJ_{ijkl}\eta_i\eta_j\eta_k\eta_l\\
H_u &=\frac{1}{8}\sum_{ijkl}^NB_{ij}B_{kl}\eta_i\eta_j\eta_k\eta_l
\end{align}
with the entries of $J_{ijkl}$ and $B_{ij}$ totally antisymmetric and drawn from independent, identically distributed gaussians with zero mean and
\begin{align}
\overline{J_{ijkl}^2} = \frac{J^2}{N^3} && \overline{B_{ij}^2} = \frac{J}{N^2}
\end{align}
where $\overline{Q}$ is our notation for the disorder average of the quantity $Q$. We also define, for our future convenience
\begin{align}
k = \frac{N}{2}&&
M = \frac{N!}{2!(N-2)!}
\end{align}
As presaged in the introduction, our approach is approximation of the thermal density matrix of each disorder realization by the thermal density matrix of a free Majorana Hamiltonian. To whit, we consider the Gibbs-Delbruck variational principle which states that if
\begin{align}
\sF = - T \ln\Tr\left[\exp\left(-\beta H\right)\right]
\end{align}
for any trial density matrix $\rho_t$
\begin{align}\label{GDVP}
\sF\leq \Tr\left[H\rho_t\right] + T\Tr\left[\rho_t\ln\rho_t\right]\equiv \sF_t
\end{align}
This suggests a standard variational approach: we minimize over some tractable class of $\rho_t$ to get the best tractable approximation to $\sF$. Notably, this also gives an approximation to the thermal density matrix of a particular disorder realization
\begin{align}
\rho = \exp\left(-\beta(H-\sF)\right)
\end{align}
since in terms of this quantity the variational principle reads
\begin{align}
0\leq  \Tr\left[\rho_t\left(\ln\rho - \ln\rho_t\right)\right]
\end{align}
and we can recognize the quantity on the right as the relative entropy, an information theoretic measure of the difference between two density matrices. Thus the optimum $\rho_t$ for the purposes of approximating the free energy is the closest to the true thermal density matrix in the sense of relative entropy.

We choose as our class of trial density matrices
\begin{align}\label{trialH}
H_t &= \frac{i}{2}\sum_{ij}^NG_{ij}\eta_i\eta_j\\
\sG &= -  T \ln\Tr\left[\exp\left(-\beta H_t\right)\right]\\\label{triald}
\rho_t &= \exp\left(-\beta(H_t - \sG)\right)
\end{align}
where $G$, our variational parameter, is any antisymmetric real matrix. Utilizing these trial density matrices should be thought of as doing mean field theory in the observable $\av{\eta_i\eta_j}$. Indeed, one can see that the equations we eventually derive for the optimum $G$ can be identified with the saddle point equations for the field generated in a Hubbard-Stratonovich decoupling of the $4$-fermion interaction.

We note for the sake of the fastidious reader that when we refer to "Gaussian random free Majoranas" we mean an ensemble of Hamiltonians of the form given in Equation~\ref{trialH} with $G_{ij}$ drawn from independent, identically distributed Gaussians with zero mean and
\begin{align}
\overline{G_{ij}^2} = \frac{J}{N}
\end{align}

Though we will find later that the parametrization of this class of density matrices by $G$ is profitable in a numerical context, we make a change of variables in order to explore the broad character of the minima. There is some $\Phi\in O(N)$ such that defining
\begin{align}
\xi_i = \sum_{j}\Phi_{ij}\eta_j
\end{align}
give new cannonical Majoranas $\xi_i$ in terms of which we have
\begin{align}
H_t = \frac{i}{2}\sum_{\mu}^{k}g_\mu \xi_{2\mu-1}\xi_{2\mu}\label{diag}
\end{align}
where the spectrum of $G$ is $\pm i g_\mu$ and we have chosen $g_\mu\geq 0$. We will refer to this $\Phi$ as "diagonalizing" $G$, since
\begin{align}
\Phi^t G\Phi = \sum_{\mu}^kg_\mu e^\mu
\end{align}
where
\begin{align}
e^{ij}_{kl} &= \delta_{i,k}\delta_{j,l} - \delta_{i,l}\delta_{j,k}\\
e^\mu & = e^{2\mu-1,2\mu}
\end{align}
Equation~\ref{diag} implies
\begin{align}
\av{\xi_{2\mu-1}\xi_{2\mu}}_t = -i\tanh\left(\frac{g_\mu}{ T}\right)\equiv -id_\mu
\end{align}
where
\begin{align}
\av{A}_t = \Tr\left[A\rho_t\right]
\end{align}
Reversing the transformation, we find
\begin{align}
\av{\eta_i\eta_j}_t = \frac{\delta_{ij}}{2} - iC_{ij}
\end{align}
with
\begin{align}
iC = \tanh\left(\frac{iG}{ T}\right)
\end{align}
where for any holomorphic function $f(z)$ and Hermitian matrix $X$ we have defined
\begin{align}\label{hol}
f(X) = \frac{1}{2\pi i }\oint_\gamma f(z)\left(z\bbI - X\right)^{-1}\dd z
\end{align}
where $\gamma$ is some contour enclosing the spectrum of $X$ and $\bbI$ is the identity matrix. 

For any $T>0$, $C(G,T)$ is a diffeomorphism from all antisymmetric real matrices to antisymmetric real matrices with eigenvales with absolute value strictly less than $1$. Since it is a diffeomorphism, we may consider our trial free energy (and density matrices) to be a function of $C$ rather than $G$. For future use, we define $\sA\cong \bbR^M$ to be the space of all antisymmetric real matrices, $\sC\subset\sA$ to be those matrices with eigenvalues with absolute value strictly less than $1$ and $\overline{\sC}\subset \sA$ to be those matrices with eigenvalues with absolute value less than or equal to $1$.

Thinking of $C$ as our variational parameter also allows us to slightly expand our class of variational density matrices in a critical way. For any $C\in \sC$, we can write
\begin{align}
C = \Phi\left(\sum_{\mu}^kd_\mu e^\mu\right)\Phi^t
\end{align}
Diagonalizing $H_t$, we can find that this $C$ maps to the density matrix
\begin{align}\label{ex}
\rho_t(C) = 2^{-k}\sum_{s\in \bbZ_2^k}\left(\prod_{\mu}^k(1+s_\mu d_\mu)\right)\ket{\Phi,s}\bra{\Phi,s}
\end{align}
where $\ket{\Phi,s}$ is the ground state of the free fermion Hamiltonian
\begin{align}
\frac{i}{2}\sum_{ij}\tilde{G}_{ij}\eta_i\eta_j && \tilde{G} = \Phi\left(\sum_{\mu}s_\mu e^\mu\right)\Phi^t
\end{align}
and we have identified $\bbZ_2$ with $(\{\pm 1\},\cdot)$. This formula can manifestly be extended to give a well defined density matrix for any $C\in\overline{\sC}$. Since $\overline{\sC}$ is compact, this guarantees that the trial free energy will attain a minimum on this class of density matrices.

This extension is not merely a curiosity, but necessary to obtain a minimum at $T = 0$. This is to be expected, since $\partial\sC = \overline{\sC} - \sC$ corresponds to density matrices of less than full rank: in particular it contains the pure states corresponding to the ground states of free Majorana Hamiltonians.

To put a fine point on it, fix $\Phi\in SO(N)$ arbitrary and consider the zero temperature minimum with respect to $d_\mu$. At $T = 0$, we have
\begin{align}
\sF_t = 2^{-k}\sum_{s\in \bbZ_2^k}\left(\prod_{\mu}^k(1+s_\mu d_\mu)\right)\inoprod{\Phi,s}{H}{\Phi,s}
\end{align}
Some $s_0\in \bbZ_2^k$ has the minimum expectation value of the energy $\inoprod{\Phi,s_0}{H}{\Phi,s_0}$ and clearly $\sF_t$ is then minimized with respect to $d$ by choosing $d=s_0$, corresponding to the pure state $\ket{\Phi,s_0}$. So, at $T = 0$, the minimum is always attained on $\partial\sC$.

\subsection{Properties of the minima}
In terms of $C$, a lengthy but straightforward calculation gives
\begin{align}\label{FEs}
\sF_t &= E - TS\\
E &= \frac{1}{2} \av{C,LC} - \frac{u}{2}\norm{B}^2\\
L &= K + uU\\
S &= N\ln(\sqrt{2}) - \frac{1}{2}\Tr\left[\left(\bbI + iC\right)\ln\left(\bbI + iC\right)\right]\label{FEe}
\end{align}
where we have defined the inner product and norm on $\sA$
\begin{align}\label{inprod}
\av{X,Y} &= \frac{1}{2}\Tr\left[XY^T\right]\\
\norm{X} &= \sqrt{\av{X,X}}
\end{align}
and $J$ and $U$ are linear functions $\sA\rightarrow\sA$ defined by
\begin{align}
\av{X,KY} &= -\frac{1}{4}\sum_{ijkl}^NJ_{ijkl}X_{ji}Y_{kl}\\
\av{X,UY} &= -\av{X,B}\av{B,Y} - \av{X,BYB}
\end{align}
We will write the latter as
\begin{align}
U = -B\otimes B - B\boxtimes B
\end{align}
Since it is actually composed of two different tensor products of $B$. $B\otimes B$ refers to the tensor product of $B$ with itself considered as a vector (in $\bbR^M$, c.f. the operator $\ket{\alpha}\bra{\alpha}$) while $B\boxtimes B$ is the tensor product of $B$ with itself considered as an operator on $\bbR^N$ projected onto the subspace of antisymmetric matrices ($\Lambda^2(\bbR^N)$).

Now, in order to characterize the minimum of the free energy as a function of $C$, we must take derivatives of $\sF_t$ with respect to $C$. This computation is lengthy and not particularly informative, so we relegate it to Appendix~\ref{deriv}. There, we find
\begin{align}\label{1cder}
\nabla_C\sF_t = L(C)+G
\end{align}
So that our minimum must have
\begin{align}
L(C) = -G\label{mincon}
\end{align}
We also have for the Hessian (the matrix of second derivatives)
\begin{align}
\hess_C(\sF_t) = L-T\hess_C(S)
\end{align}
As for $-\hess_C(S)$, we give a more thorough characterization in Appendix~\ref{deriv} and for now simply note that its eigenvalues are given by
\begin{align}\label{HSEV}
\frac{g_\mu\pm g_\nu}{d_\mu\pm d_\nu}\text{ and } \frac{1}{1-d_\mu^2}
\end{align}
Using the concavity of $(1+x)\ln (1+x) + (1-x)\ln(1-x)$ on $(-1,1)$, we can see that the former is bounded below by the latter. The latter is then easily seen to be bounded below by $1$, which implies that $-(\bbI+\hess_C(S))$ is non-negative. So, if $\lambda_m$ is the minimum eigenvalue of $L$, for $T>\abs{\lambda_m}$ $\sF_t$ is convex in $C$. Since $\overline{\sC}$ is convex, $\sF_t$ must have a unique minimum. Since one can easily see that $\left.\nabla_C\sF_t\right|_{C = 0} = 0$ this minimum occurs at $C = 0$, giving $G = 0$ and $\rho_t = \bbI$.

For $T<\abs{\lambda_m}$, though, the extremum at the origin becomes a saddle point. To whit,
\begin{align}
\left.\hess_C(\sF_t)\right|_{C = 0} = L+ T\bbI
\end{align}
so that at $T = \lambda_m$ the minimum eigenvalue of the hessian passes through zero and the minimum moves off of the origin, stabilized by higher order terms in $S$. For $T<\abs{\lambda_m}$, there are actually a pair of minima away from the origin related by time reversal symmetry. That is, in the language of mean field theory, the individual disorder realization breaks time reversal symmetry ($G\rightarrow -G$) at the mean field level. We do not have an argument that these local minima remain the global minima below the transition temperature, and will simply ignore the question in this work. 

\subsection{Order Parameter}
\label{op}
We adopt the convention that a quantity $Q$ evaluated at the minimum will be denoted $Q_*$. $\norm{C_*}^2$ is an order parameter for the time reversal symmetry breaking, so it is natural to ask what implication it has for possible glassy order. $\norm{C_*}^2$ is in some sense a natural analogue of the Edwards-Anderson order parameter\cite{EASG}, since 
\begin{align}
\norm{C_*}^2 = -\sum_{ij}\av{\eta_i\eta_j}_{t*}\av{\eta_i\eta_j}_{t*}
\end{align}
or, in the language of replicas
\begin{align}\label{replicas}
\overline{\norm{C_*}^2} = -\sum_{ij}\av{\eta_i^\alpha\eta_j^\alpha\eta_i^\beta\eta_j^\beta}
\end{align}
Thus, the presence of nonzero $\overline{\norm{C_*}^2}$ in the thermodynamic limit is our signal that the physics is governed by a non replica-diagonal saddle point (i.e. "spin glass" order) and a lack of self averaging in $\norm{C_*}^2$ is our signal for replica symmetry breaking.\cite{RSBREV} 

We note in passing that $\overline{\norm{C_*}^2}$ contains all non-trivial information from the first two disorder moments of the distribution of $\norm{C_*}^2$, due to the $O(N)$ statistical symmetry of the SYK model. It is trivial to see that this symmetry forces $\overline{C_*} = 0$. However, it also gives that
\begin{align}
\overline{C_{ij*}C_{kl*}} = \left(\delta_{ik}\delta_{jl} -\delta_{il}\delta_{kj}\right)\frac{2\overline{\norm{C_*}^2}}{N(N-1)} 
\end{align}
by applying Schur's Lemma to the adjoint representation of $SO(N)$. While the fundamental reflections in $O(N)$ allow us to see that the third disorder moment must be zero, the fourth disorder moment is in general allowed to be more complicated. In this work, we study only $\overline{\norm{C_*}^4}$ as a test of self-averaging in $\norm{C_*}$.

To forestall questions of whether these correlations are sub-leading, we note that
\begin{align}
0\leq -\frac{1}{2}\sum_{i\neq j} \av{\eta_i\eta_j}\av{\eta_i\eta_j} \leq \frac{N}{2}
\end{align}
for any density matrix, not just a thermal density matrix of a free Majorana Hamiltonian. To see this, notice that
\begin{align}
M_{ij} = \av{\eta_i\eta_j} - \frac{\delta_{ij}}{2}
\end{align}
is an antisymmetric Hermitian matrix by the properties of the Majorana operators. So there must be $\tilde{\Phi}\in O(N)$ such that
\begin{align}
M = \tilde{\Phi}\left(\sum_\mu^kim_\mu e^\mu\right)\tilde{\Phi}^t
\end{align}
and defining the new cannonical Majoranas
\begin{align}
\tilde{\xi}_i = \sum_{ij}^N\tilde{\Phi}_{ij}\eta_j
\end{align}
we find
\begin{align}
\abs{m_\mu} = \abs{\av{\tilde{\xi}_{2\mu-1}\tilde{\xi}_{2\mu}}}\leq 1
\end{align}
implying
\begin{align}
-\frac{1}{2}\sum_{i\neq j} \av{\eta_i\eta_j}\av{\eta_i\eta_j} = \sum_{\mu}^km_\mu^2\leq \frac{N}{2}
\end{align}

Since the minimum is attained by a pure state at $T = 0$, we have
\begin{align}\label{opscal}
\left.\norm{C_*}^2\right|_{T = 0} = \frac{N}{2}
\end{align}
for every disorder realization. So, this implies that
\begin{align}
\left.\overline{\norm{C_*}^2}\right|_{T = 0} = \frac{N}{2}
\end{align}
and $\overline{\norm{C_*}^2}$ saturates this bound at $T=0$.

\subsection{Large $N$ scaling}
\label{Nscal}
Beyond this point, making further analytic progress seems daunting, since we must exert some understanding of the random symmetric real matrix $L$. $L$ is not drawn from any well studied matrix ensemble we are aware of, rendering hope of this understanding slim. Some partial progress and modest intuition can be developed about $L$'s constituent parts, however, which can then be checked numerically.

We begin with $U$, since the distribution of $B$ is relatively well understood. First, we can notice that $B\otimes B$ is rank one with nonzero eigenvalue
\begin{align}
\lambda_\otimes = \norm{B}^2
\end{align}
It is a trivial application of the central limit theorem to see that
\begin{align}\label{otscale}
\sqrt{\frac{N(N-1)}{3J^2}}\left(\frac{N^2}{N(N-1)}\lambda_{\otimes} - \frac{J}{2}\right)\xrightarrow{d} N(0,1)
\end{align}
where $N(\mu,\sigma)$ denotes a normally distributed random variable with mean $\mu$ and variance $\sigma^2$, so that in the large $N$ limit, $\lambda_\otimes$ becomes narrowly distributed around $J/2$.

Our understanding of $B\boxtimes B$ is more limited, but some can still be said without straining. Since the eigenvalues of the tensor product of two operators are the product of the eigenvalues of the operators, we find the bound for the operator norm
\begin{align}
\norm{B\boxtimes B}_{op}\leq \norm{B}_{op}^2
\end{align}
Since $iB$ is drawn from a Hermitian Wigner matrix ensemble, we have\cite{taotopics}
\begin{align}
\norm{B}_{op} = O\left(\frac{1}{\sqrt{N}}\right)
\end{align}
and so
\begin{align}
\norm{B\boxtimes B}_{op}= O\left(\frac{1}{N}\right)
\end{align}

Before synthesizing these results, we say what we can about $K$. Unfortunately, this amounts entirely to conjecture that will be borne out in our numerical results. $K$ bears some resemblance to a sample from a symmetric Wigner matrix ensemble, except for the fact that
\begin{align}
\av{e^{ij},Ke^{ik}} = 0
\end{align}
for every $i$, $j$ and $k$. This violates the identical distribution assumption and prevents us from rigorously applying any of those results. Heuristically, however, we can notice that there are $\nCr{M}{2}$ off diagonal elements and only $\nCr{N}{3}$ of them are correlated. So, the proportion of them that are correlated is
\begin{align}
\frac{\nCr{N}{3}}{\nCr{M}{2}} = O\left(\frac{1}{N}\right)
\end{align}
and it is perhaps reasonable to expect that we have
\begin{align}
\norm{K}_{op} = O\left(\frac{1}{\sqrt{N}}\right)
\end{align}
as we would were $K$ actually drawn from a symmetric Wigner distribution.

As to what this means for our minimization problem, we have the simple eigenvalue bound
\begin{align}
\lambda_m \geq -\norm{K}_{op} - u\norm{B\boxtimes B}_{op}- u\Theta(u)\lambda_\otimes
\end{align}
giving
\begin{align}\label{lamscal}
\lambda_m = \left\{\begin{array}{lr}O(1) & u>0\\O\left(\frac{1}{\sqrt{N}}\right) & u\leq 0\end{array}\right.
\end{align}
provided that our conjecture about the scaling of $\lambda_K$ is correct.

Assuming that where it is nonzero $\norm{C_*}^2 = O(N)$ (as is suggested by Equation~\ref{opscal}), we can use Equation~\ref{lamscal} to deduce the scaling of a few other quantities. Notably, we find
\begin{align}
E_* = \left\{\begin{array}{lr}O(N) & u>0\\ O(\sqrt{N}) & u\leq 0\end{array}\right.
\end{align}
which casts serious doubt on how well this approximation captures the $T = 0$ physics of the model for $u\leq 0$, since we expect an extensive (i.e. $O(N)$) ground state energy for all $u$. This question will be discussed in more detail in Section~\ref{dis}. Finally, we notice that the minimum condition allows us to see
\begin{align}
\norm{G_*}^2 =\left\{\begin{array}{lr}O(N) & u>0\\ O(1) & u\leq 0\end{array}\right.
\end{align}

\subsection{The Susceptibility and Heat Capacity}
\label{singav}
We also compute an approximation to the susceptibility and heat capacity within this framework. The details of this computation are given in Appendix~\ref{2der}. There, we find that the susceptibility shows a singularity of the form
\begin{align}
\chi_0 \sim \frac{\text{sgn}(\abs{\lambda_m} - T)}{\abs{\lambda_m} - T}
\end{align}
We can compute a crude approximation to the average of the susceptibility by simply integrating this expression against the distribution of the lowest eigenvalue $\lambda_m$:
\begin{align}
\overline{\chi_0} \sim \int_{-\infty}^\infty\frac{\text{sgn}(\abs{\lambda_m} - T)}{\abs{\lambda_m} - T}p(\lambda_m)\dd \lambda_m
\end{align}
which, combined with our expectation that $p(\lambda_m)$ should be supported on all of $(0,\infty)$ for finite $N$, yields an integral that does not exist. That is, we can conclude that the disorder average for this approximation to the susceptibility does not exist at finite $N$.

Heuristically, one should perhaps think about this result in the reverse direction. At any given temperature, there's a finite probability of finding the transition temperature of a disorder realization within any given small region around the temperature of interest. Given the strength of the singularity in the susceptibility, these nearby transitions have large enough values of the susceptibility to prevent the average from converging at the temperature of interest. It is perhaps illustrative to consider circumstances under which this could fail to happen. If the singularities predicted in an individual disorder realization were less severe, say $\abs{T-\abs{\lambda_m}}^{-1/2}$, this integral would converge and our heuristic would predict finite disorder averages. There, we would be saved by the fact that we do not have finite probability of finding the transition \textit{precisely} at the temperature of interest and the nearby transitions are not strongly enough singular to make up for this fact.

We note in passing that this behavior is certainly an artifact of our variational approximation. For finite $N$, the true free energy is analytic in the probe field defined in Appendix~\ref{2der}. We should perhaps have expected worse behavior out of the susceptibility than our other quantities of interest, since it is not controlled directly by a variational principle like our other quantities of interest (with the exception of the heat capacity). We also note that its possible that a prediction of the susceptibility in this framework might be salvageable with more careful analysis, since $\lambda_m$ self-averages at large $N$ which appears to superficially address the issue. A careful examination of this approach and whether it can be accessed numerically is outside the scope of the present work, however, and we make no further inquiry into the susceptibility.

As for the heat capacity, the expression given in Equation~\ref{heatcap} bears some investigation for any dangerous singularities given our experience with a similar expression for the susceptibility. However, considering a Landau-type expansion in small $G$ near $T=\abs{\lambda_m}$ leaves one with the expectation that 
\begin{align}
\norm{G_*} = O\left(\left(\abs{\lambda_m}-T\right)^{1/2}\right)
\end{align}
as $T\rightarrow\abs{\lambda_m}^-$ so that counting powers of $\abs{\lambda_m}-T$ then suggests that $C_V=O(1)$ in this limit. Numerically, we see no evidence of a singularity in the heat capacity in an individual disorder realization, though note a discontinuity in the heat capacity at $T=\abs{\lambda_m}$.

\section{Numerical Results}
\label{numset}

\begin{figure}
\includegraphics[scale=0.53]{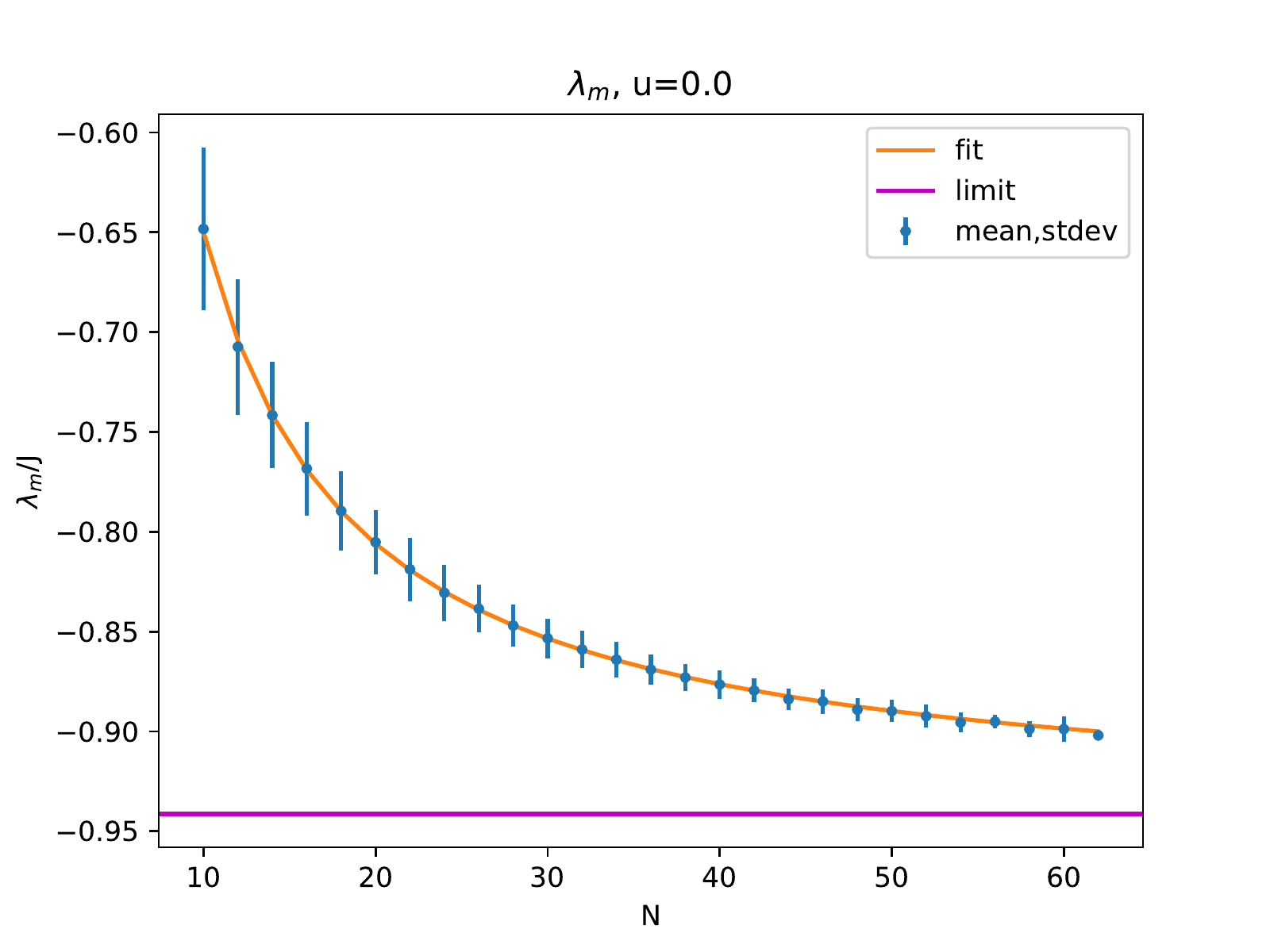}
\caption{\label{lam0}Averages and mean squared errors for $\lambda_m$ for $u = 0$, together with fit line and limiting value.}
\end{figure}

To complement our analysis above, we numerically find the minima of randomly generated samples of $L$ to characterize the statistics of various quantities at the minimum. All of our code was written in Python $2.7.14$ using SciPy $1.0.0$ and NumPy $1.14.0$ and run on the CNSI "Knot" cluster at UCSB. After a sample of $L$ is generated, its minimum eigenvalue and the associated eigenvector are found using SciPy's eigh function. The code then searches for a minimum of $\sF_t$ as a function of $G$ using BFGS as implemented in SciPy's minimize function, starting with the temperature just below the minimum eigenvalue and with an initial guess just away from the origin in the direction of the minimum eigenvector of $L$. After a minimum is found at a given temperature, the code decreases the temperature by a predetermined step and searches for the minimum starting at the previous minimum. An exact value for the gradient with respect to $G$ is provided, as calculated in Appendix~\ref{deriv}. We choose to minimize with respect to $G$ rather than $C$ to avoid having to numerically enforce the eigenvalue constraint. We of course study the quantities $\sF_{t*}$, $E_{*}$, $S_{*}$ and $C_V$. To track some information about the minimizing state and the resulting distribution, we also track $\norm{C_*}^2$ and $\norm{G_*}^2$.

\begin{figure}
\includegraphics[scale=0.53]{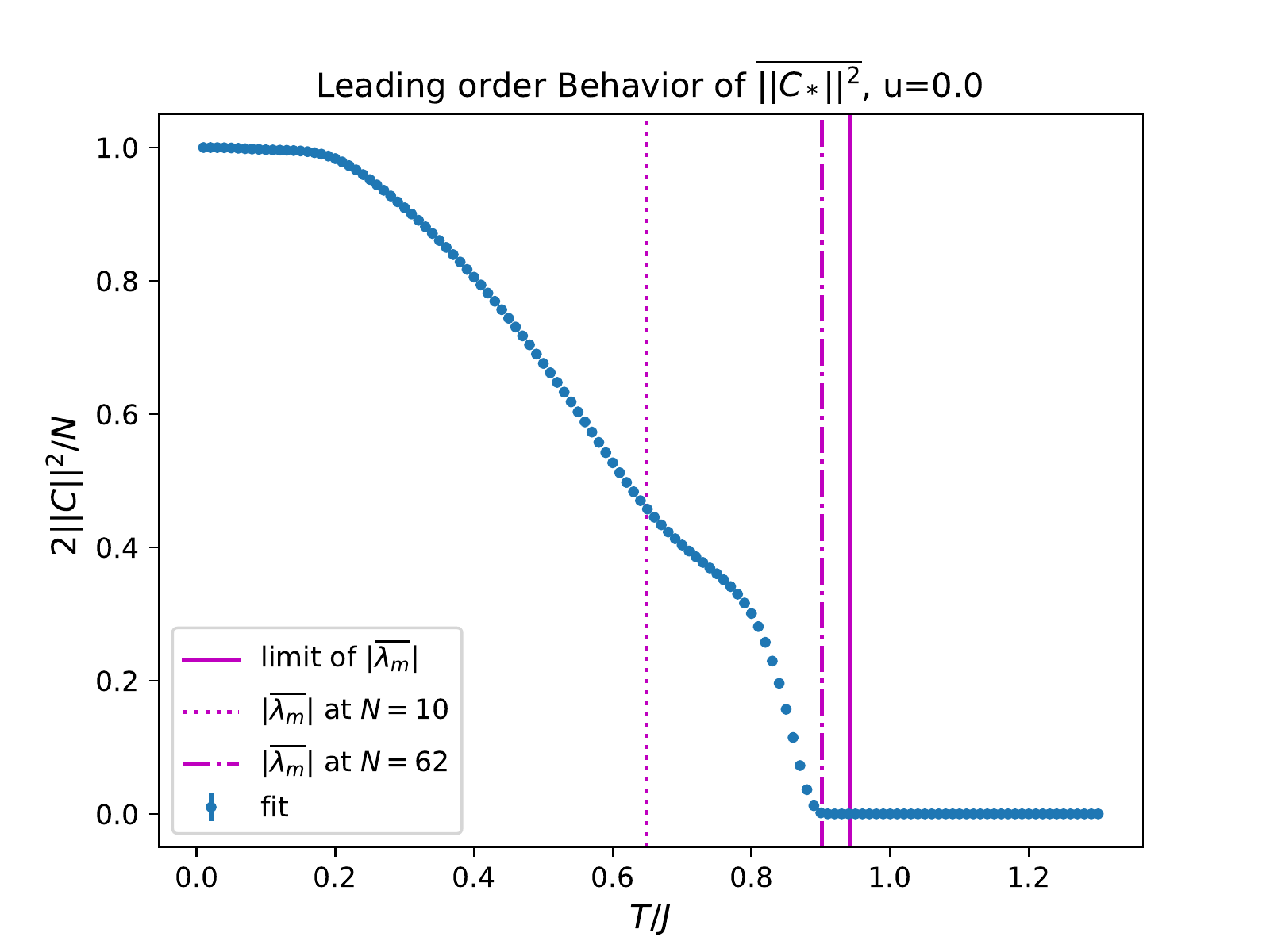}
\caption{\label{c0}Coefficient of $N$ in $\norm{C_*}^2$ regression for $u=0$, with error bars.}
\end{figure}

To counteract the fact that we expect $\overline{\lambda}_m\rightarrow 0$ for $u\leq 0$, we rescale the model and study instead
\begin{align}
\tilde{L} = \left\{\begin{array}{lr}2L & u>0\\\frac{3\sqrt{N}}{4}L & u \leq 0\end{array}\right.
\end{align}
This rescales all quantities with dimensions of energy identically, so $\tilde{T}$ and $\tilde{G}$ are rescaled by the same factors. The factors of two exist to move the limiting value of $\overline{\lambda}_m$ close to $J$ as observed empirically. All $N$ dependencies given in this section will be given in terms of the rescaled model, rather than the original model. Because of our limited computational resources, we study only three values of $u$, $u = \pm 1$ and $u = 0$. These values were chosen with the expectation from Bi et al's work and our analysis so far that there are only two relevant regions of $u$, $u\leq 0$ and $u>0$ together with a modest degree of hedging that the $u = 0$ case might conceivably be special.

\begin{figure}
\includegraphics[scale=0.53]{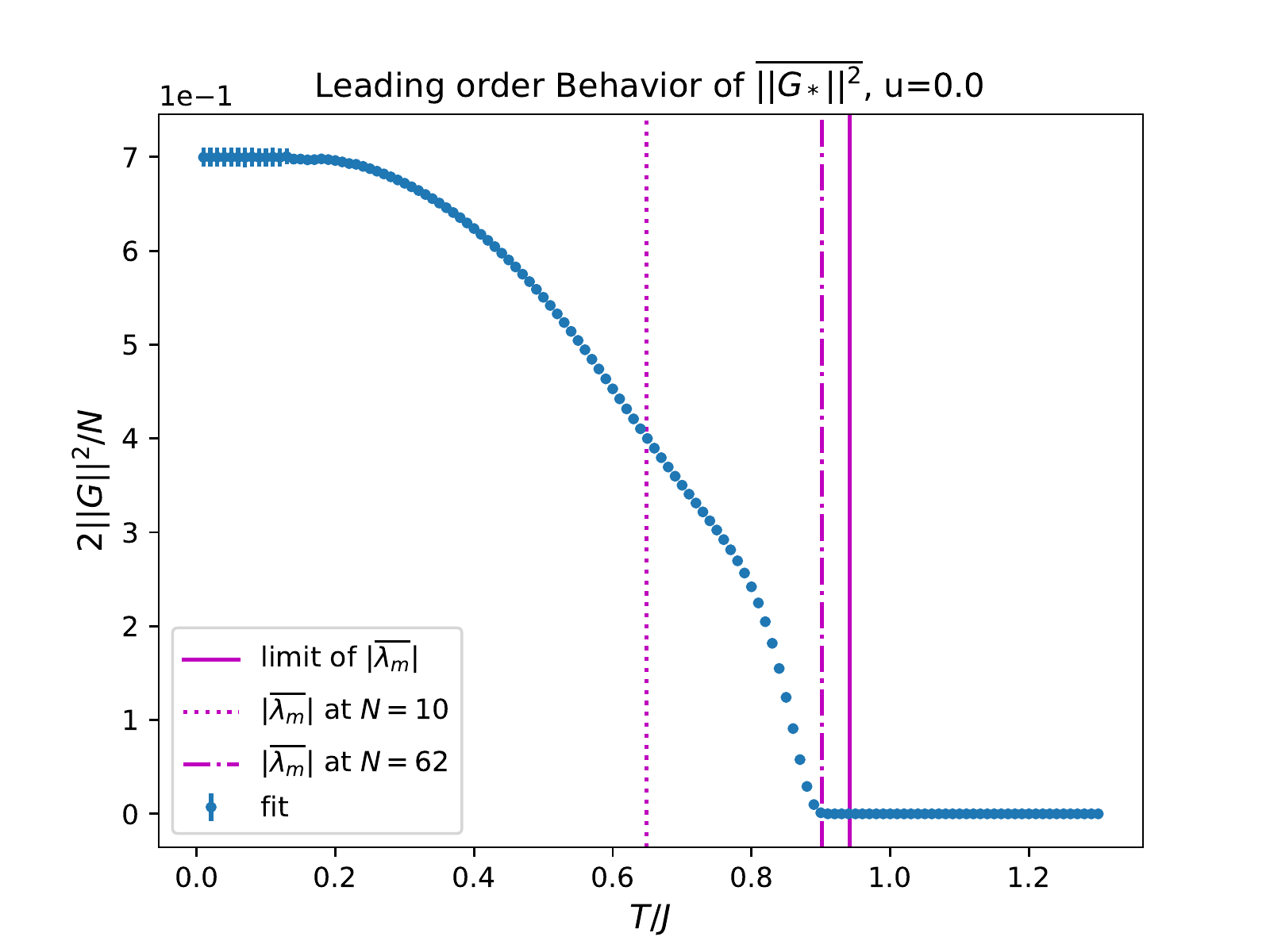}
\caption{\label{g0}Coefficient of $N$ in $\norm{G_*}^2$ regression for $u=0$, with error bars.}
\end{figure}

We are interested in the leading order large $N$ behavior of our quantities of interest as a function of temperature. However, only $N \leq 62$ is numerically accessible to us without considerable effort. Our numerical efforts produced $500$ samples for $N = 10$ through $N = 30$ and a steadily decreasing number of samples through $N=62$ where we received only $12$ samples. We sampled every available $N$ in this range, i.e. every even $N$. Consequently, we must work just a little bit to extract information about the large $N$ limit with our available data. Our analysis largely follows White\cite{het-fit} with some trivial modifications, but we present the techniques here to ensure we are clear about what we mean by each quantity and in case the reader is unfamiliar. We will be concerned with two questions for all of our quantities of interest: what is the leading order behavior of their average and do they exhibit self-averaging.

\begin{figure}
\includegraphics[scale=0.53]{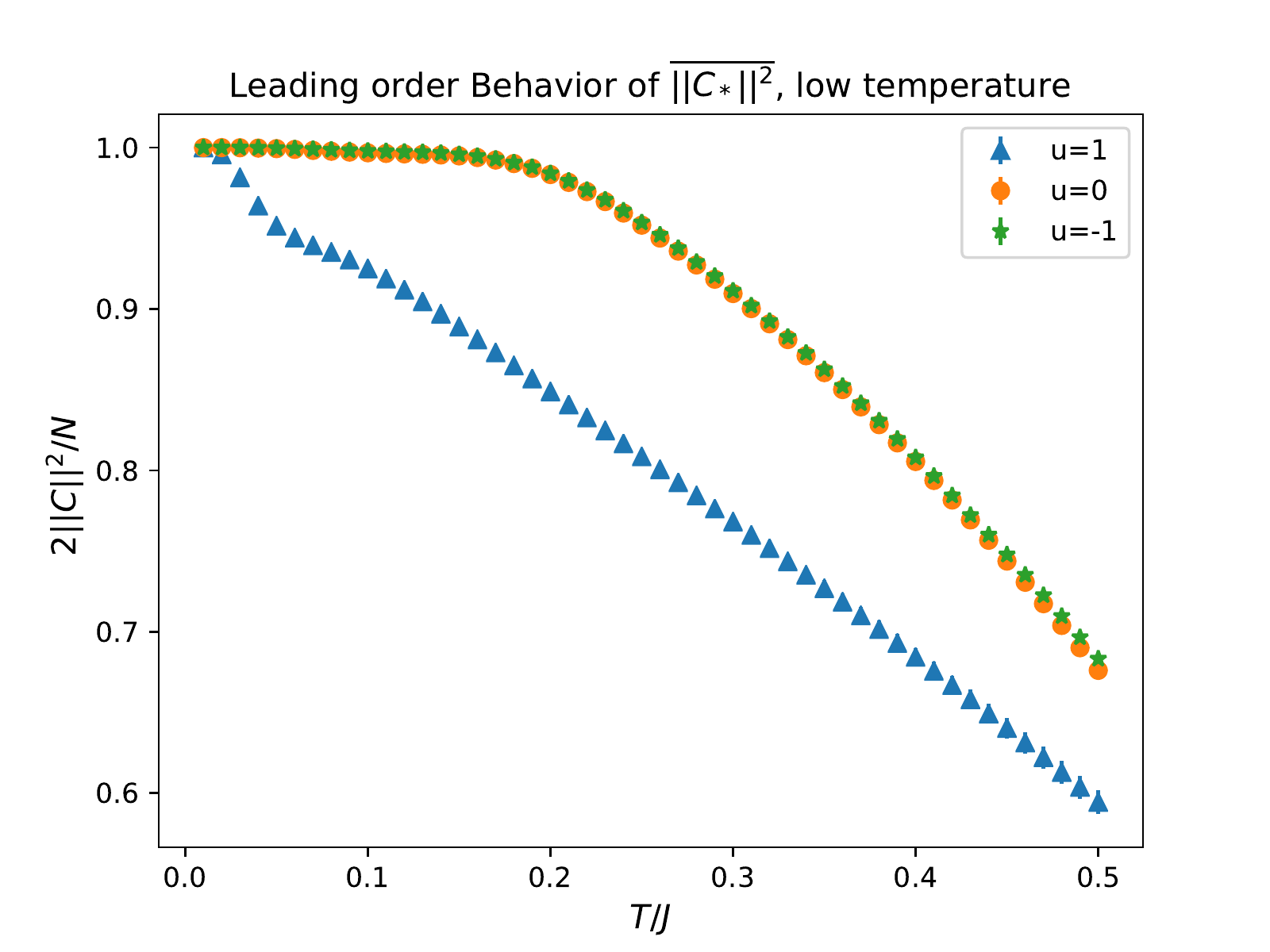}
\caption{\label{inset}Coefficient of $N$ in $\norm{C_*}^2$ for various $u$ at low temperature.}
\end{figure}

For the first question, consider some quantity $y = O(N^\nu)$. Largely we will be concerned with quantities with $\nu = 1$, with the exception of $\lambda_m$ which has $\nu=0$. We will attempt to fit our observations to a model of the form
\begin{align}
y_i = \sum_{j =0}^{2}N_{i}^{\nu-j} y_{f,j} + \epsilon_i
\end{align}
where $i$ stands for the $i$th of $n$ independent observations at various $N$ and $\epsilon_i = y_i - \overline{y}_i$. As is standard for regression problems, we phrase this in terms of matrices as
\begin{align}
Y = \sN Y_f + \epsilon
\end{align}
where $Y$ is now a $n\times 1$ vector of observations, $\sN$ is our $n\times 3$ "design matrix" which contains the powers of $N$ we expect the averages to depend on and $Y_f$ is a $3\times 1$ vector of our unknown fit parameters.

\begin{figure}
\includegraphics[scale=0.53]{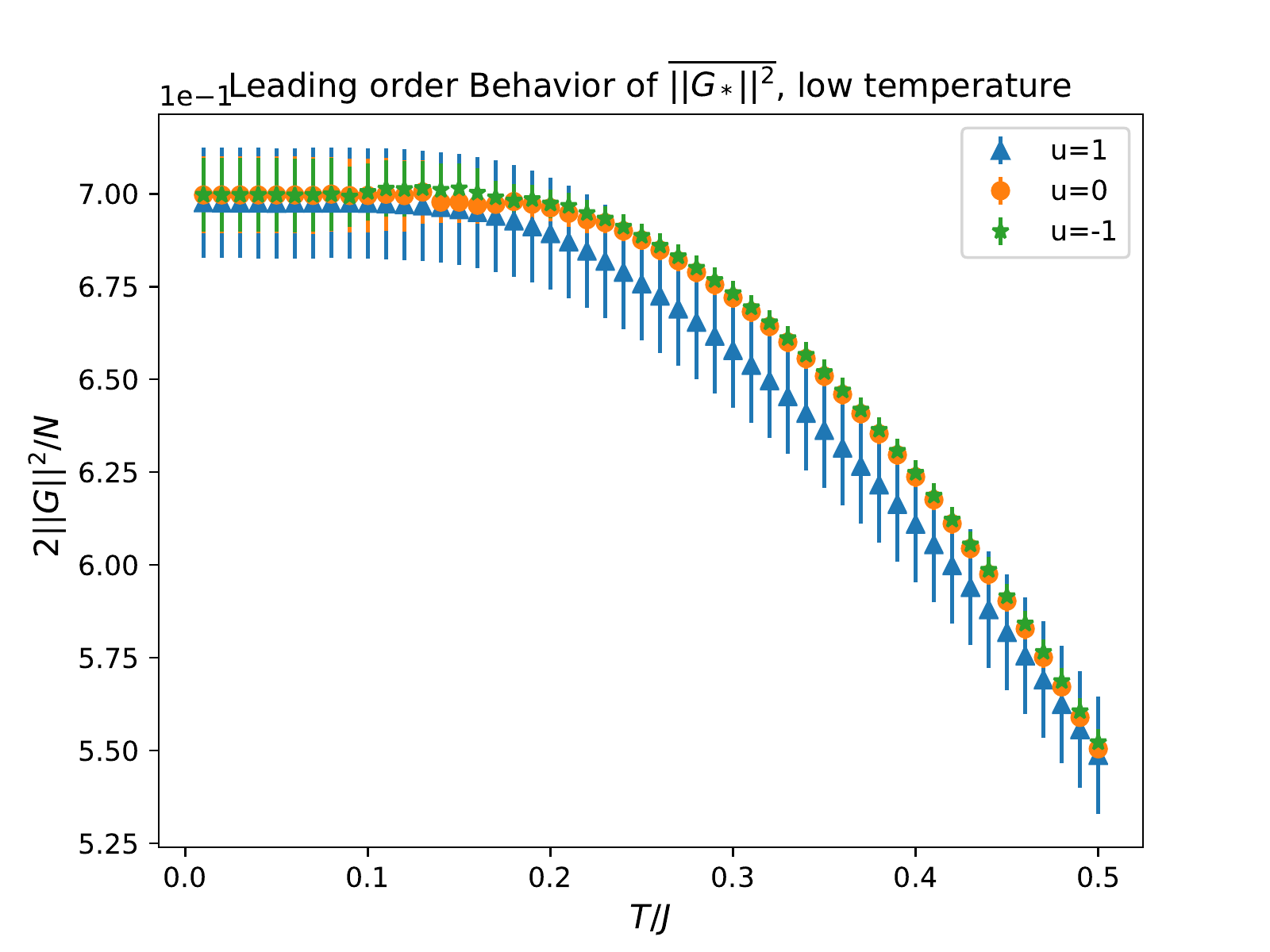}
\caption{\label{insetg}Coefficient of $N$ in $\norm{G_*}^2$ for various $u$ at low temperature.}
\end{figure}

The critical difference between this situation and a standard regression problem is that the variances of the $\epsilon_i$ are both unknown and expected not to be equal. Even so, if we define the ordinary least squares estimator
\begin{align}
\hat{Y}_f = \left(\sN^t\sN\right)^{-1}\sN^tY
\end{align}
and the variance estimator
\begin{align}
\hat{V} &= \left(\sN^t\sN\right)^{-1}\sN^t\hat{R}\sN\left(\sN^t\sN\right)^{-1}\\
\hat{R}_{ij} &= \delta_{ij}\left(y_i - \left[\sN\hat{Y}_f\right]_i\right)^2
\end{align}
White\cite{het-fit} shows that as the number of observations grows
\begin{align}\label{fitasym}
\hat{V}^{-1/2}\left(\hat{Y}_f - Y_f\right)\xrightarrow{d} N(0,\bbI_3)
\end{align}
Since one can observe that $\hat{V}=O(n^{-1})$, where $n$ is the number of observations, this shows that $\hat{Y}_f$ is still a consistent estimator for our fit parameters. This also allows us to compute confidence intervals for these parameters (albeit only asymptotically correct ones). All confidence intervals quoted will be the $99\%$ confidence windows under the asymptotic distribution. We have not accounted for multiple testing in our statistical procedure. Treated carefully, this could be expected to widen our confidence intervals by up to a factor of $5$, depending on how exactly we treated the asymptotic distribution in Equation~\ref{fitasym}. This would not materially impact the conclusions presented below.

\begin{figure}
\includegraphics[scale=0.53]{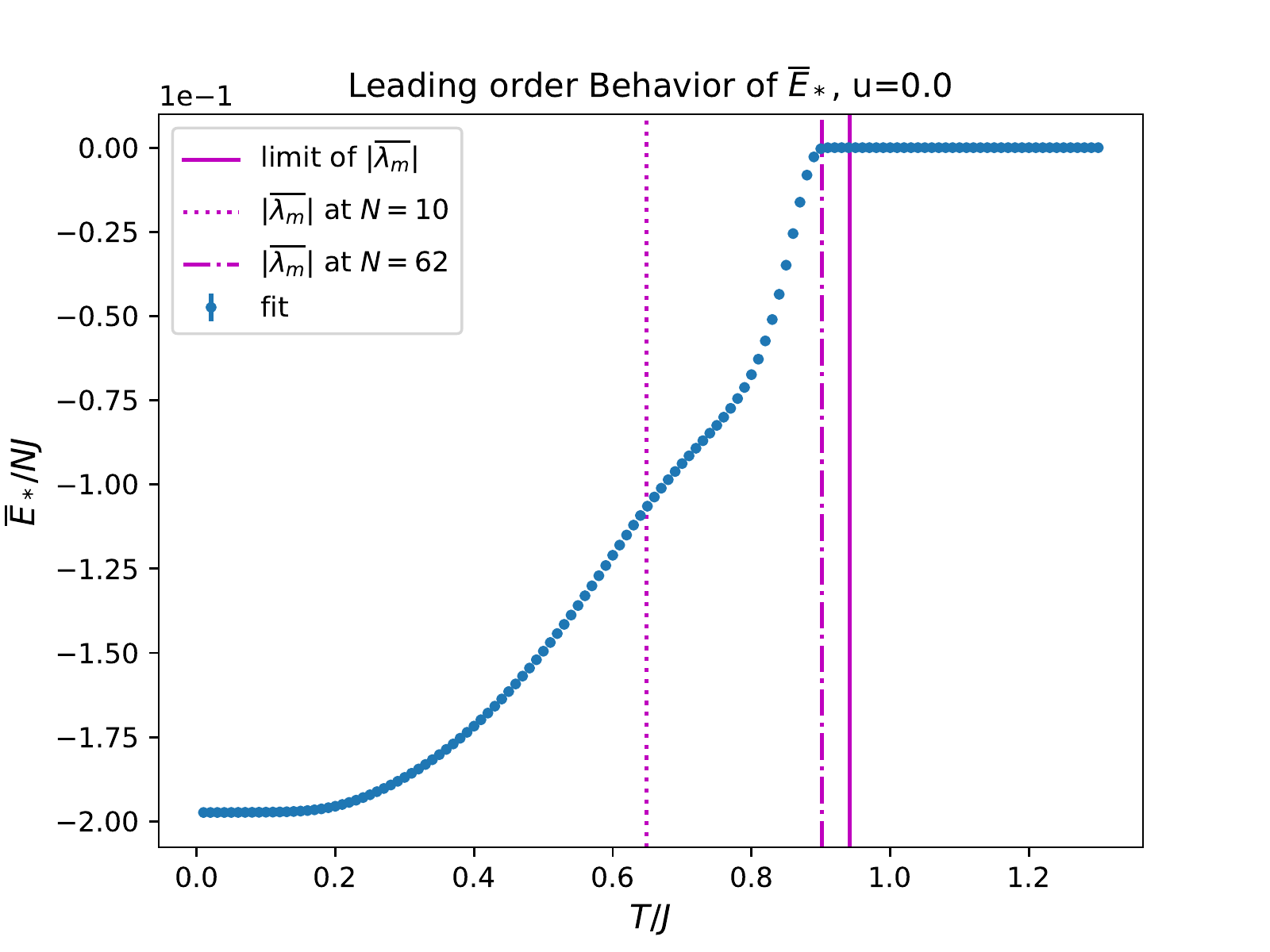}
\caption{\label{e0}Coefficient of $N$ in $E_*$ regression for $u=0$, with error bars.}
\end{figure}

The question of self-averaging requires a modicum more work, since we must be careful about what exactly constitutes an observation of the variance. One could imagine a number of ways to organize this information, but we simply use all data points taken at a particular $N$ to construct an estimate of the variance at that $N$ and count this as a single observation of the variance. Since $y$ undergoes self averaging if $y^2 - \overline{y}^2 = O(N^{2\nu-1})$ rather than $O(N^{2\nu})$, we probe self-averaging by fitting our observed values of the variance to $N^{2\nu}$ through $N^{2\nu-3}$ and reporting the leading order coefficient with $99\%$ confidence windows.

\begin{figure}
\includegraphics[scale=0.53]{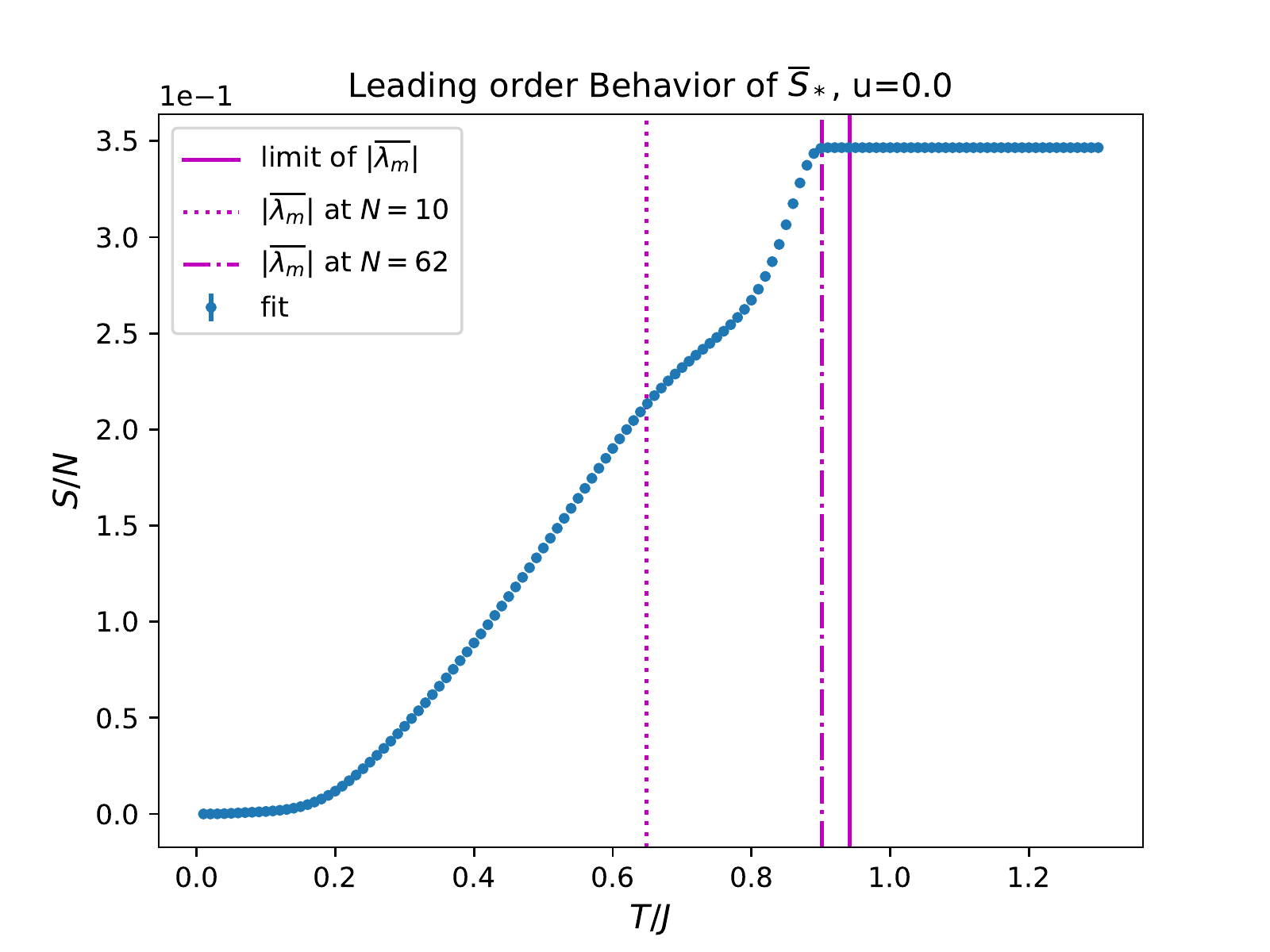}
\caption{\label{s0}Coefficient of $N$ in $S_*$ regression for $u =0$, with error bars.}
\end{figure}

By this point, there is relatively little in the numerical results that is a surprise, since we have already conjectured all of our highest leverage results. The scalings conjectured in Section~\ref{Nscal} are consistent with our observed scalings in the rescaled model. Comfortingly, we find that the fit to the data for $\lambda_m$ gives for $u = 1$ that $\lambda_m\rightarrow 0.99\pm 0.01$, which is consistent with the limiting value one would expect from assuming $\lambda_\otimes$ is the dominant contribution to the large $N$ limit and using Equation~\ref{otscale}. In particular, our results strongly argue that our scaling expectations for $\norm{C_*}^2$ (FIG.~\ref{c0}) are correct, an indication of glassy behavior in the model. We also note the consistency of the numerical results of these quantities with many of their known $T = 0$ limiting values.

\begin{figure}
\includegraphics[scale=0.53]{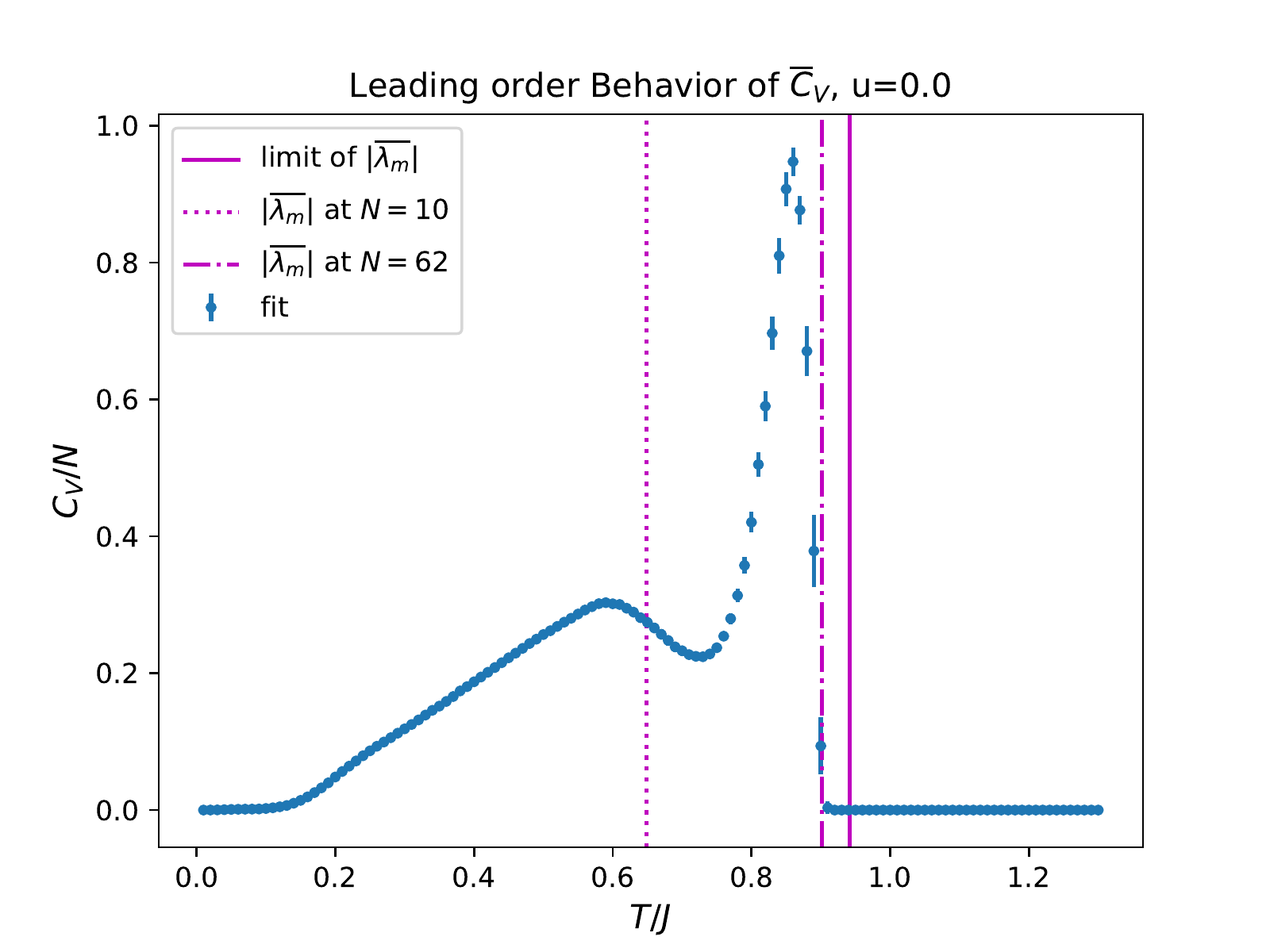}
\caption{\label{cv0}Coefficient of $N$ in $C_V$ regression for $u =0$, with error bars.}
\end{figure}

Looking at our results as a whole, two broad trends bear discussion first. The first of these themes is that the results for $u = -1$ and $u =0$ look nigh identical to the human eye. One might have expected this given the information on $B\boxtimes B$ presented in Section~\ref{Nscal}, since this term is expected to be sub-leading relative to $K$. One might also have expected this on the basis that previous study of this model indicates $u\leq 0$ should all be a single phase.\cite{bi_xu} A close inspection of the data for $\norm{C_*}$ at $u=1$, however, reveals a subtle feature at low temperature (FIG.~\ref{inset}) that is hard to conclusively make sense of with the available data.  We conjecture that this is due to differences in the angular distribution (or eigenvalue distribution, if the reader prefers) of $G_*$, about which more will be said in Section~\ref{dis}. This conjecture is supported by the presence of this feature in our data for $S_*$ and $C_V$ along with the lack of any such feature in the data for $\norm{G_*}^2$ and is consistent with the lack of this feature in the data for $E_*$ and $\sF_{t*}$.

\begin{figure}
\includegraphics[scale=0.53]{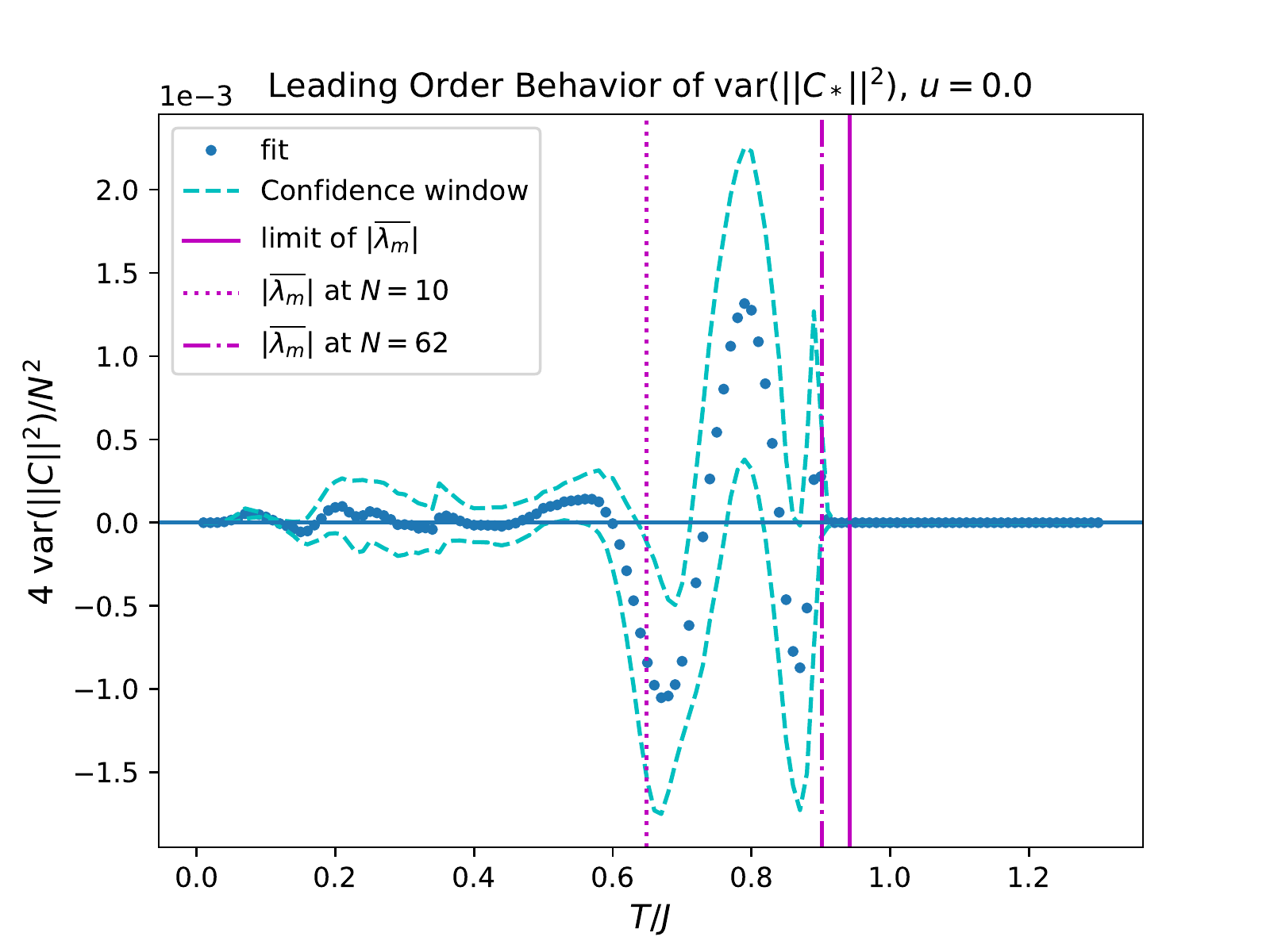}
\caption{\label{vc0}Coefficient of $N^2$ in of $\norm{C_*}^2$ regression for $u =0$, with confidence window.}
\end{figure}

On the second theme: we note a few regions in the plots which almost certainly show finite size effects. The most clear instance of this is the lack of an $O(N)$ component in the temperature range between the limiting value of $\abs{\overline{\lambda_m}}$ and the value of $\abs{\overline{\lambda_m}}$ at our final $N$ point, $N=62$. We have a strong expectation that this region of temperatures should actually be in the ordered phase, but samples of disorder realizations with $\abs{\lambda_m}$ greater than the average value at $N=62$ are quite rare at all of the $N$ points that we sample. This is perhaps more intuitive upon looking at our fits of the observed $\lambda_m$ (FIG.~\ref{lam0}) where one can notice that $\lambda_m$ appears to self average more rapidly than it converges to its limiting value. This accounting is supported by the fact that the plots for $u=1$ show this feature much less strongly while $\lambda_m$ appears to self average much more slowly for $u=1$. For similar reasons, we regard any dramatic features in the fits in the region between the values of $\abs{\overline{\lambda_m}}$ for $N=10$ and $N=62$ with a mild suspicion, as temperatures further to the right of that region spend progressively more of our sample artificially above the transition temperature.

\begin{figure}
\includegraphics[scale=0.53]{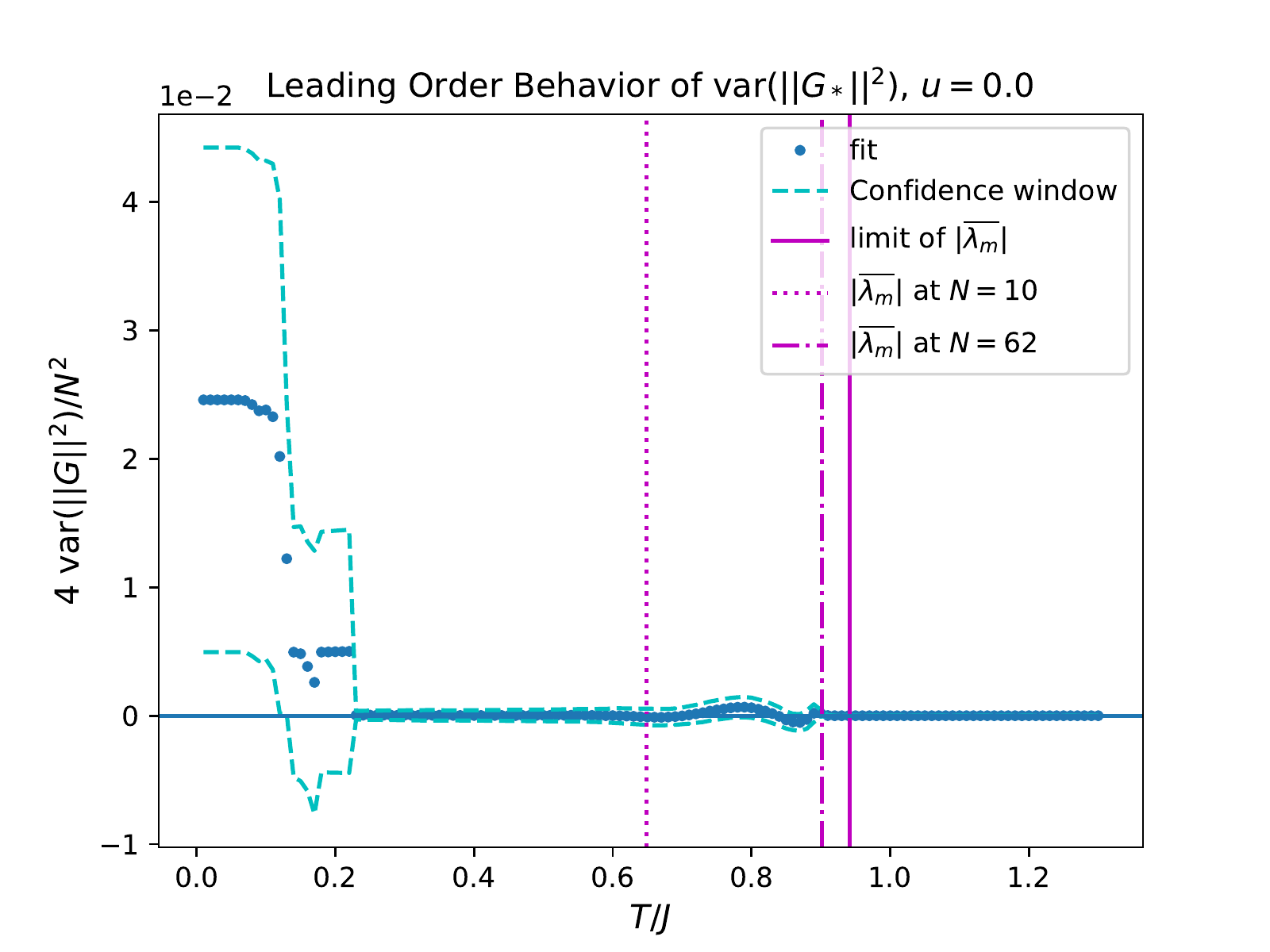}
\caption{\label{vg0}Coefficient of $N^2$ in of $\norm{G_*}^2$ regression for $u =0$, with confidence window.}
\end{figure}

One might reasonably rouse some suspicion towards our statistical analysis on these grounds, since it does not raise any red flags in the form of wider confidence intervals in most of our quantities in these regions. However, ultimately this is not so surprising since these finite size effects represent "unknown unknowns" from the point of view of the statistical techniques. Upon being presented with a large number of data points which largely cluster around a zero slope line, there is no statistical basis to expect that the line might suddenly upturn at a later data point or that our knowledge of the slope is likely to be imprecise. This is the problem of induction, not a problem with our analysis.

\begin{figure}
\includegraphics[scale=0.53]{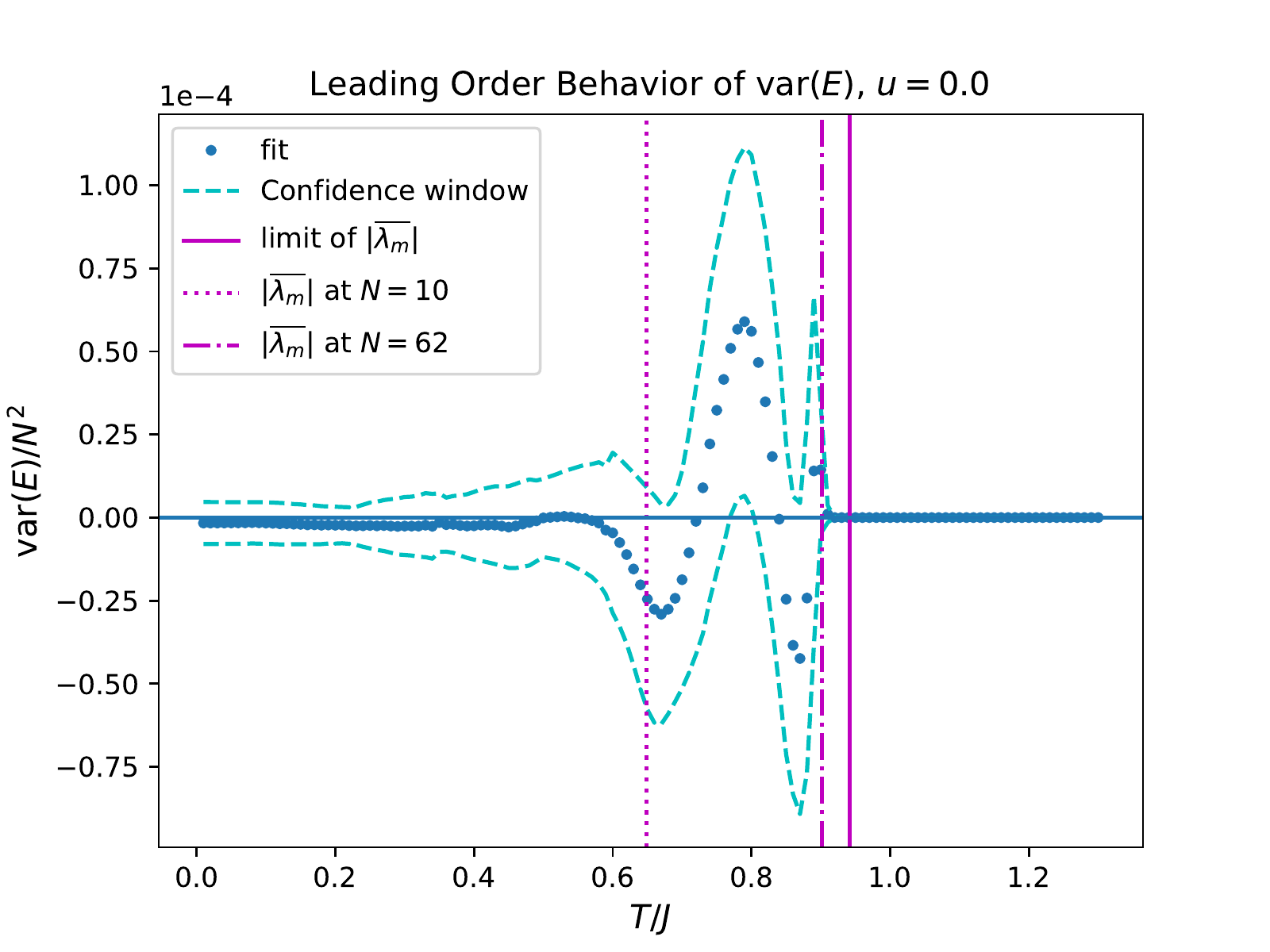}
\caption{\label{ve0}Coefficient of $N^2$ in of $E_*$ regression for $u = 0$, with confidence window.}
\end{figure}

Finally, before we move on to questions of self-averaging, we briefly discuss the heat capacity (FIG.~\ref{cv0}). Unfortunately, our data is inconclusive as to whether the average heat capacity will develop a singularity at the limiting value of $\abs{\overline{\lambda_m}}$ or simply reproduce the discontinuity seen in individual disorder instances. This question is ultimately governed by the $N$ scaling of the $T\rightarrow \abs{\lambda_m}^-$ limit of the heat capacity of each individual disorder realization. We find this question rather inscrutable based only on Equation~\ref{heatcap} at present. This question might be within the scope of additional numerical attacks of this problem, with higher $N$ and a finer gradation of temperature points.

\begin{figure}
\includegraphics[scale=0.53]{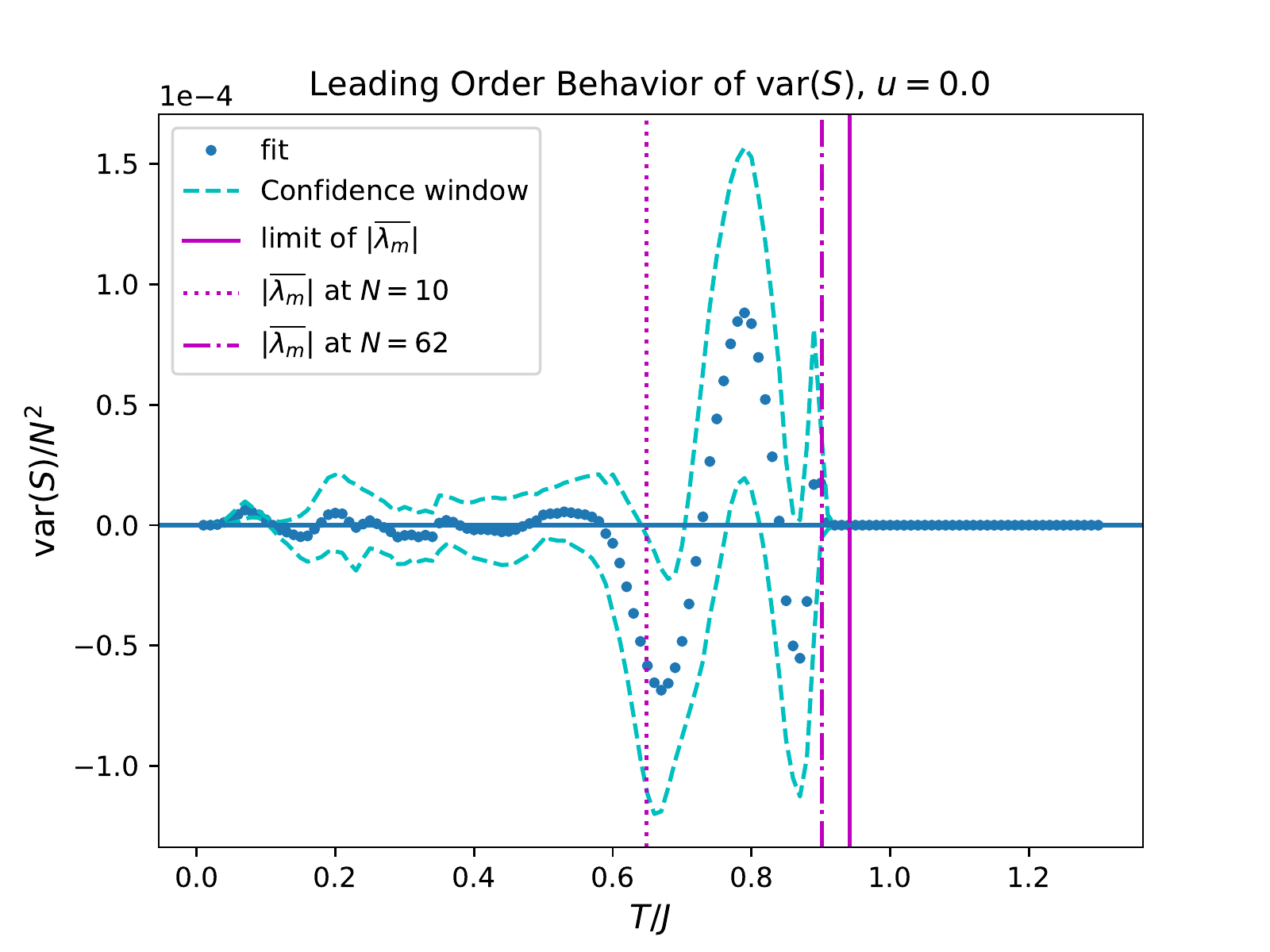}
\caption{\label{vs0}Coefficient of $N^2$ in of $S_*$ regression for $u = 0$, with confidence window.}
\end{figure}

Our ability to make conclusive statements about self averaging is somewhat weaker than our ability to address questions about the averages themselves, unfortunately. We lay the blame here on the slower convergence of the variance estimates of all of our quantities than that of the estimates of their mean. Fortunately, the behavior we do see in the variance estimates looks unambiguously more like noise than signal. Largely, our standard confidence interval includes $0$, meaning that we can not reject the hypothesis that these variances are zero with $99\%$ confidence. Even taking into account the fact that some regions of the plots \textit{do} put $0$ outside of this confidence interval and that a more modest confidence interval (e.g. $95\%$) would widen these regions, we still do not find compelling evidence for a lack of self averaging in any of these quantities. Two considerations lead us to this conclusion. Firstly, we find that what estimates we do have for some non-zero $O(N^2)$ component of the variance of some quantity are largely orders of magnitude lower than the $O(N^2)$ component of the square of the mean of this quantity, laying some of the blame at the feet of our inability to detect the full cancellation of two large numbers. Secondly, the behavior of these quantities where they are largest is in many cases inconsistent with our physical or statistical expectations. These estimates are occasionally negative, which is impossible for any true leading-order contribution to a variance. Much of the action in these plots is concentrated in the region near the transition, which we have already flagged as a likely haven of finite size effects. Many other regions of concern (e.g. low temperature in FIG.~\ref{vg0}) occur at temperatures where the fits of the averages show sudden spikes in the size of the confidence window, indicating that these might be driven by errors or outliers.	

\begin{figure}
\includegraphics[scale=0.53]{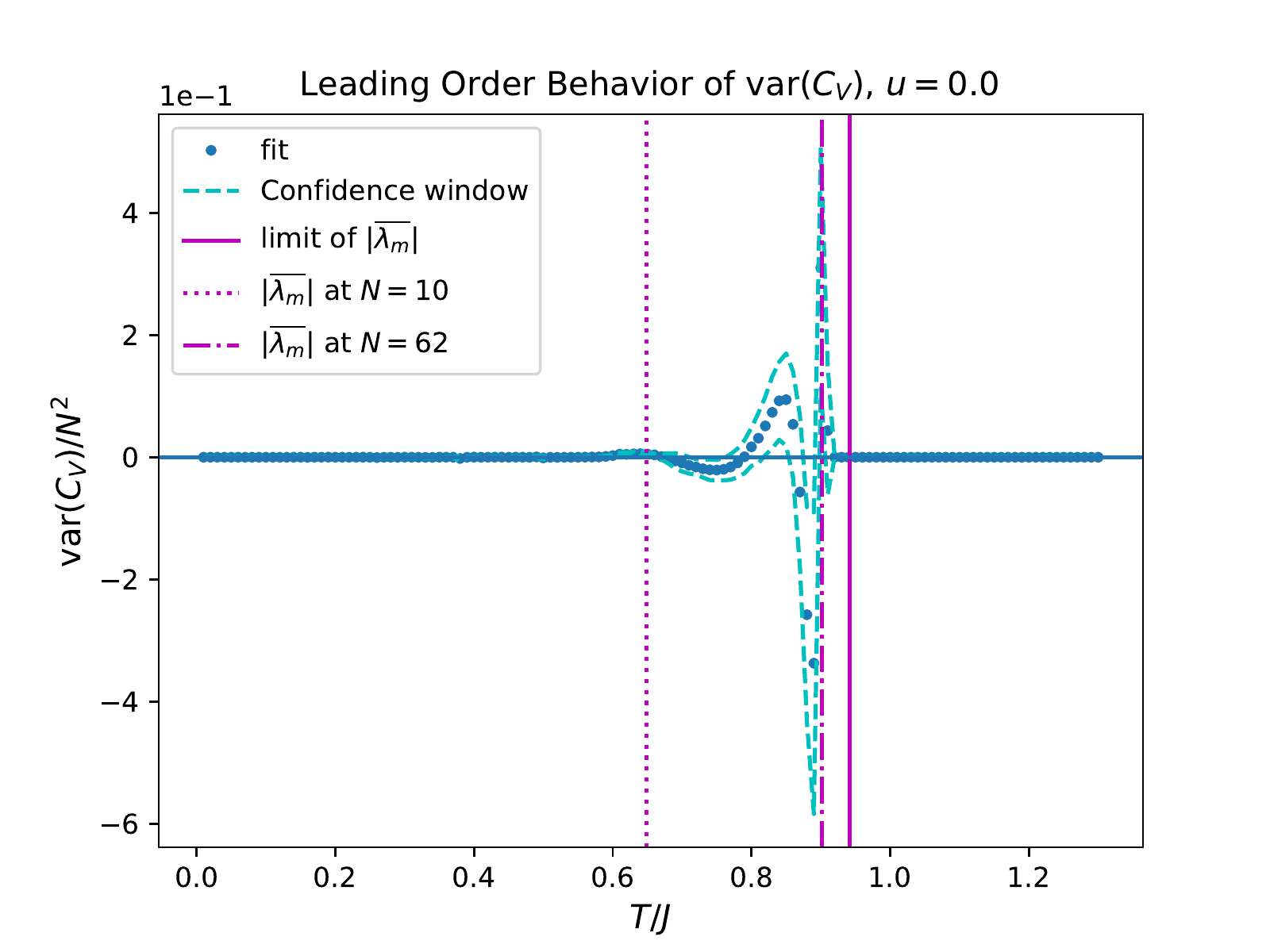}
\caption{\label{vcv0}Coefficient of $N^2$ in of $C_V$ regression for $u = 0$, with confidence window.}
\end{figure}

\section{Discussion}
\label{dis}
For the $u>0$ phase, we have reason to expect that this approximation captures the properties of the time reversal symmetry breaking phase quite well. In fact, as $u\rightarrow \infty$ and $T = 0$, the true ground state becomes arbitrarily close to a state in our class of variational state since (as noted by Bi et al\cite{bi_xu})
\begin{align}
H_u = -\frac{1}{2}\left(\frac{i}{2}\sum_{ij}B_{ij}\eta_i\eta_j\right)^2
\end{align}
has a two-fold degenerate ground state spanned by a pair of free Majorana ground states. Indeed, in this limit we can see that $G_*\rightarrow \pm f(N)B$ so we expect that the distribution of $G_*$ ought to behave like that of Gaussian random free Majoranas up to some scaling. This expectation is borne out in what numerical results we have, though as we noted we cannot actually fully characterize the distribution of $G_*$ with the moments we have studied. One might express some surprise that $G_* = O(N)$ while $B = O(1)$ (in the sense of norms). However, this is the scaling that one finds in the Gaussian random free Majoranas (set to give an extensive free energy), so this scaling is necessary to prevent the distribution of density matrices from approaching infinite or zero temperature in the $N\rightarrow \infty$ limit.

We take a moment to reconcile what might be an apparent difference between our results and those of Bi et al\cite{bi_xu}, the presence of replica non-diagonal terms in this model. While they find that they can ignore replica indexes in their analysis, they also proceed by considering the boson
\begin{align}
b = \frac{i}{2}\sum_{ij}B_{ij}\eta_i\eta_j
\end{align}
which the above argument suggests should behave like $\av{G_*,C_*}$ in its disorder statistics, up to scaling. Using the minimum condition, we can actually identify this quantity with $E_*$, which self-averages around its first moment according to our analysis. So, the boson should indeed be replica diagonal at the saddle point, as they observed. 

Assessing the implications for the replica index structure of the fermion Green's function is somewhat harder, however. One might recall Equation~\ref{replicas} and be tempted to assume the saddle point value of the fermion Green's function is simply not actually replica diagonal. However, this configuration corresponds to a phase that breaks fermion parity, i.e. one that has
\begin{align}
\overline{\av{\eta_i}^2}\neq 0
\end{align}
which is inconsistent with our observations in the present work and existing observations by Bi et al. Returning to our analogy with the Sherrington-Kirkpatrick model, we expect that any glassy order suggested here should be captured in the replica off-diagonal structure in the fermion four point function. In Bi et al's work, this is completely determined by the saddle point value of the two point function due ultimately to their choice of Hubbard-Stratonovich decoupling. As outlined above, the two point function alone cannot produce replica off-diagonal physics without breaking fermion parity. Thus, in order to capture this physics in the field theoretic framework, one must re-examine how the $8$-fermion interaction in the disorder-averaged replicated partition function is decoupled	.

For $u\leq 0$, these results are more a call to action than a conclusive accounting. Since $\lambda_m\rightarrow 0$ in this region of parameter space, we are left with the prediction of a phase transition at $T =0$ into a phase that breaks time reversal symmetry and exhibits some glassy behavior. The particulars of this story should be viewed with a healthy dose of skepticism, since as noted in Section~\ref{Nscal} the variational ground state energy is sub-extensive ($O(\sqrt{N})$) as $N\rightarrow \infty$. In particular, this precludes the variational ground state from having an $O(1)$ overlap with the true ground state in this limit.

As an aside, this exposes of a curiosity of this analysis that is perhaps worthy of discussion in its own right: free fermion states are remarkably poor at capturing the physics in this model outside of the $u>0$ ordered state despite the "average freedom" noted in Section~\ref{intro}. One aspect of this is the low overlap of any free fermion ground state with the true ground state at $u\leq0$ noted above. To the extent that one is willing to interpret an SYK ground state as a generic ground state of a Hamiltonian composed only of interactions, this shows that such states are usually orthogonal to free fermion ground states. We can also see that in the high temperature phase (i.e. the phase with the emergent conformal symmetry) $\rho_{t*}$ self-averages around $\rho_{t*} = 2^{-N/2}\bbI$. That is, in the large $N$ limit, the variational states become certain that they can say precisely nothing about the ensemble of thermal density matrices. One could perhaps argue in their favor that they get the average right, since one can show that an $O(N)$ statistical symmetry of the form possessed by this model forces $\overline{\rho} = 2^{-N/2}\bbI$. Given that the ensemble of $\rho_{t*}$ must also possess this $O(N)$ symmetry, however, this doesn't argue in their favor above any other $O(N)$ symmetric class of trial states.

Within the context of this analysis, we expect that while the $T = 0$, $u\leq 0$ phase may break time reversal and exhibit some glassy behavior, it is unlikely to be the same phase as the low temperature $u>0$ phase. One piece of evidence is actually the change in scaling of the ground state energy, since this is a rather dramatic change between these two regions of parameter space. We also notice that two $O(N)$ symmetric distributions for $G$ with $\overline{\norm{G}^4} = (\overline{\norm{G}^2})^2$ need not have the same distribution, in contrast with the case for an $O(N)$ symmetric distribution for an $O(N)$ vector. This is related to the fact mentioned in Section~\ref{op} that the $O(N)$ symmetry begins to fail to constrain the moments of $G$ (or $C$) to a single scalar starting at the fourth moment. Ultimately, this is due to the fact that the action of $O(N)$ cannot affect the eigenvalues of $G$ beyond permutation. 

With this in mind, reviewing the absence of the low temperature feature seen in $\norm{C_*}^2$ (FIG.~\ref{inset}) for $u=1$ from the graphs for $u = 0$ and $u = -1$ suggests quite strongly that we are seeing distinct distributions of eigenvalues in the $T\rightarrow 0$ limit of $G_*$. Since the graphs of $\norm{G_*}^2$ (FIG.~\ref{insetg}) all appear roughly identical and to have converged to their $T\rightarrow 0$ limit by the time the feature is present in $u=1$, this feature cannot be due to shifts in the overall size of $G_*$ as a function of temperature. Rather, it suggests that there are some eigenvalues of $G_*$ in the $u = 1$ case that are typically lower than in the $u =0 $ and $u = -1$ case and so are "frozen out" only at lower temperatures. We should be cautious, however, about interpreting this difference as certainly indicating a difference between the $N\rightarrow \infty$ limits of these distributions, however, since there are also finite size differences between the two cases. Notably: the $u=1$ case self-averages more slowly, due to the smaller number of components of $B$ relative to $J$.

In sum, our results represent a tantalizing glimpse into the low temperature physics of the SYK model. We hope that they spark further investigations of the low temperature physics of this model and inform explorations of the non-analyticity predicted at $T = 0$ by the replica calculation.

\begin{acknowledgments}
We would like to thank Cenke Xu, Leon Balents, Tim Hsieh, Kelly Pawlak and Yi-Zhuang You for helpful discussions. We acknowledge support from the Center for Scientific Computing from the CNSI, MRL: an NSF MRSEC (DMR-1720256) and NSF CNS-1725797.
\end{acknowledgments}

\appendix

\section{Derivatives of the Trial Free Energy}
\label{deriv}
Our strategy for computing the necessary derivatives will be extending these functions from functions of antisymmetric matrices (thought of as purely imaginary Hermitian matrices) to all Hermitian matrices, using this extension to diagonalize the argument to ease our computation and then restricting the resulting derivatives to act only on antisymmetric matrices. We set our notation to ease determining the domain of any formulas. While we use the notations of $\nabla$ and $\hess$ for the first two derivatives of functions on $\sA$ to match the notions on $\bbR^M$, we will denote the directional derivative of the function $f(X)$ of a Hermitian matrix $X$ at a point $X$ in the directions $\{B_i\}$ by
\begin{align}
D^nf(X;\{B_i\}) = \left.\left(\prod_{i}^n\frac{\partial }{\partial \epsilon_i}\right)f\left(X+\sum_{i}^n\epsilon_i B_i\right)\right|_{\epsilon_i=0}
\end{align}
where $D^nf(X;\{B_i\})$ of course takes values in the same space that $f$ does (typically, for our purposes, $\bbR$ or hermitian matrices.) In this context, $Df$ is our notation for the gradient of a scalar function (using the modification of the standard inner product on Hermitian matrices which restricts to the inner product given in Equation~\ref{inprod}):
\begin{align} 
\frac{1}{2}\Tr\left[Df(X)Y^\dagger\right] = Df(X;Y)
\end{align}
If $f$ takes values among Hermitian matrices instead, $Df$ will refer to its Jacobian
\begin{align}
Df(Y)Z = Df(Y;Z)
\end{align}
Similarly, we write $D^2f$ to mean the Hessian of a scalar function. That is, $D^2f = D(Df)$.

There is a natural action of $\Theta\in U(N)$ on all Hermitian matrices by
\begin{align}
R(\Theta)(X) = \Theta X \Theta^\dagger
\end{align}
Many of our functions will be invariant under this action, so we notice that if $f(R(\Theta)X) = f(X)$ then we have
\begin{align}\label{rot}
D^nf(R(\Theta)X;\{B_i\}) = D^nf(X;\{R(\Theta^\dagger)B_i\})
\end{align}

Our final ingredient will be a method for taking derivatives of functions defined by Equation~\ref{hol}. For this, we notice
\begin{align}\label{basic}
D^1 X^{-1}(X;B) = -X^{-1}BX^{-1}
\end{align}
by using the product rule and linearity of scalar derivatives applied to the equation $XX^{-1} = \bbI$. Applying this to a function of the form given in Equation~\ref{hol} gives
\begin{align}
D^1f(X;B) = \frac{1}{2\pi i }\oint_\gamma f(z)\left(z\bbI - X\right)^{-1}B\left(z\bbI - X\right)^{-1}\dd z
\end{align} 

We actually begin with derivatives of $\sG$, for reasons which will gradually become clear. Using standard manipulations on free fermion Hamiltonians, we find
\begin{align}
\sG = -\frac{ T}{2}\left(N\ln(2) + \Tr\left[\ln\cosh\left(\frac{iG}{T}\right)\right]\right)
\end{align}
Applying the chain rule, we can actually compute directly with the above technique that for antisymmetric $B$
\begin{align}
D^1\sG(iG;iB) = -\av{B,C(G)}
\end{align}
or
\begin{align}\label{1gder}
\nabla_G\sG = -C
\end{align}
We can also compute from our knowledge of free fermions that
\begin{align}
\av{H_t}_t = -\av{G,C}
\end{align}
This gives
\begin{align}
TS = \av{G,D_G\sG}-\sG
\end{align}
and, since we will see in a second that $\sG$ is concave in $G$, $TS$ is the Legendre transform of $\sG$. Hence,
\begin{align}
\nabla_C(-TS) = G
\end{align}
justifying the non-trivial portion of Equation~\ref{1cder}. We also have
\begin{align}
T\hess_C(S) = \hess_G(\sG)^{-1}
\end{align}
allowing us to finish all the derivatives with respect to $C$ that we need once we compute $\hess_G(\sG)$. Utilizing Equation~\ref{rot}, we find
\begin{align}
\hess_G(\sG) = \tilde{R}(\Phi)P_AR(E)D^2\sG(\tilde{G})R(E^\dagger)P_A^t\tilde{R}(\Phi^t)
\end{align}
where $P_A$ is the projection from Hermitian matrices to antisymmetric Hermitian matrices,
\begin{align}
\tilde{G} &= \bigoplus_\mu^k \begin{pmatrix}g_\mu & 0 \\0 & -g_\mu\end{pmatrix}\\
E &= \frac{1}{\sqrt{2}}\bigoplus_\mu^k\begin{pmatrix}1 & 1\\-i & i\end{pmatrix}\\
\tilde{R}(\Phi) &= P_AR(\Phi)P_A^t
\end{align}
and $\Phi$ "diagonalizes" $G$ in the sense of Section~\ref{Setup} so that
\begin{align}
iG = R(\Phi)R(E)\tilde{G}
\end{align}
On a practical level, we only give formulas for
\begin{align}
H = P_AR(E)D^2\sG(\tilde{G})R(E^\dagger)P_A^t
\end{align}
and then recognize that
\begin{align}
\tilde{R}(\Phi) = \Phi\boxtimes\Phi
\end{align}
which is sufficient for all of our numerical purposes. Since $\tilde{R}(\Phi)$ is orthogonal, this also allows us to fully characterize the eigenvalues of $\hess_G(\sG)$, which is sufficient for all of our analytic arguments.

As for $H$, a computation using Equation~\ref{hol} and Equation~\ref{basic} gives that
\begin{align}
D^2\sG(\tilde{G};X,Y) &= -\frac{1}{2}\sum_{ij}X_{ij}Y_{ji}\frac{w_i-w_j}{h_i-h_j}\\
w_i & = (-1)^{i+1}d_{\left\lceil \frac{i}{2}\right\rceil}\\
h_i &= (-1)^{i+1}g_{\left\lceil \frac{i}{2}\right\rceil}\\
\left\lceil \frac{i}{2}\right\rceil &= \left\{\begin{array}{lr}\frac{i}{2}&i\in 2\bbZ\\\frac{i+1}{2}&\text{else}\end{array}\right.
\end{align}
After an unpleasant calculation, we can use this to find for $\mu\neq \nu$ and $\epsilon_i =0$ or $1$
\begin{widetext}
\begin{align}
\av{e^{2\mu-1+\epsilon_1,2\nu-1+\epsilon_2},He^{2\mu-1+\epsilon_1,2\nu-1+\epsilon_2}} &= -D_{\mu\nu}^{0}-D_{\mu\nu}^1\\
\av{e^{2\mu-1+\epsilon_1,2\nu-1+\epsilon_2},He^{2\mu-\epsilon_1,2\nu-\epsilon_2})} &= -(-1)^{\epsilon_1+\epsilon_2}(D_{\mu\nu}^{0}-D_{\mu\nu}^1)\\
\av{e^{2\mu-1,2\mu},He^{2\mu-1,2\mu}}& = -\frac{1}{T\cosh^2\left(\frac{g_\mu}{ T}\right)}
\end{align}
\end{widetext}
where
\begin{align}
D_{\mu\nu}^\epsilon=\frac{\tanh\left(\frac{g_\mu}{T}\right)-(-1)^\epsilon\tanh\left(\frac{g_\nu}{T}\right)}{g_\mu-(-1)^\epsilon g_\nu} 
\end{align}
and any matrix element left unmentioned is $0$. We can see that $e^{\mu}$ is an eigenvector of $H$ with eigenvalue
\begin{align}
-\frac{1}{T\cosh^2\left(\frac{g_\mu}{ T}\right)} = -\frac{1}{T}\frac{1}{1-d_\mu^2}
\end{align}
The remaining non-zero matrix elements can be seen by inspection to be block diagonal in the $2\times 2$ blocks
\begin{align}
\begin{pmatrix}
-D_{\mu\nu}^{0}-D_{\mu\nu}^1 & \pm D_{\mu\nu}^{0} \mp D_{\mu\nu}^1\\\pm D_{\mu\nu}^{0} \mp D_{\mu\nu}^1 & -D_{\mu\nu}^{0}-D_{\mu\nu}^1
\end{pmatrix}
\end{align}
which have eigenvalues $-D_{\mu\nu}^0$ and $-D_{\mu\nu}^1$, justifying Equation~\ref{HSEV} and implicitly completing all analysis of the derivatives of the trial free energy with respect to $C$.

We need a few modest results about derivatives with respect to $G$ for Section~\ref{numset} and Appendix~\ref{2der} which we give now. Using the chain rule, Equation~\ref{1gder} and Equation~\ref{1cder} we find
\begin{align}\label{tgder}
\nabla_G\sF_t  = -\hess_G(\sG)\left(L(C)+G\right)
\end{align}
which is sufficient for our numerical needs. In the next section, we will also make use of the fact that
\begin{widetext}
\begin{align}
\label{hessG}
\av{X,\text{Hess}_G(\sF_t)Y} = \av{X,\left(\text{Hess}_G(\sG)L\text{Hess}_G(\sG)-\text{Hess}_G(\sG)\right) Y}- D^3\sG\left(iG;iX,iY,i(L(C)+G)\right)
\end{align}
\end{widetext}
which can be obtained by differentiating Equation~\ref{tgder} and making use of the product rule where applicable.

\section{Specific Heat and Susceptibility Calculations}
\label{2der}
For the heat capacity, we have
\begin{align}
C_V = \frac{\partial E_*}{\partial_T} = \av{LC_*,\frac{\partial C_*}{\partial T}}
\end{align}
Using the chain rule and Equation~\ref{1gder} gives
\begin{align}
\frac{\partial C_*}{\partial T}=- \text{Hess}_G(\sG)_*\left(\frac{\partial G_*}{\partial T} - \frac{G_*}{T}\right)
\end{align}
while differentiating the minimum condition (Equation~\ref{mincon}) gives
\begin{align}
L\left(\frac{\partial C_*}{\partial T}\right) = -\frac{\partial G_*}{\partial T}
\end{align}
Putting these together, we have
\begin{align}\label{heatcap}
C_V = \av{G_*,\frac{1}{T}\text{Hess}(\sG)_*\left(L\text{Hess}(\sG)_*-\bbI\right)^{-1}G_*}
\end{align}

For the susceptibility, we add a probe field
\begin{align}
H_h = -\frac{i}{2}\sum_{ij}h_{ij}\eta_i\eta_j
\end{align}
and take two derivatives
\begin{align}
\chi = \left.\hess_{h}\sF_{t*}\right|_{h = 0}
\end{align}
Since $\chi$ is a linear operator $\sA\rightarrow\sA$, its disorder average will be a multiple of the identity by Schur's lemma:
\begin{align}
\overline{\chi} = \overline{\chi_0}\bbI
\end{align}
where we have defined
\begin{align}
\chi_0 = \frac{1}{M}\Tr\left[\chi\right]
\end{align}

Taking one derivative, we see by the chain rule
\begin{align}
\nabla_h\sF_{t*} = C_* + \left(D_hG_*\right)^t\nabla_G\sF_{t*} + \left(D_hC_*\right)^th
\end{align}
where $D_h$ refers to the Jacobian. Using the minimum condition gives, comfortingly,	
\begin{align}
\left.\nabla_h\sF_{t*}\right|_{h = 0} = C_*
\end{align}
Using the minimum condition  and chain rule again gives
\begin{widetext}
\begin{align}\label{Hhess}
\left.\hess_h\sF_{t*}\right|_{h = 0} = \left(D_hG_*\right)^t\hess_G\left(\sF_{t}\right)_*D_hG_* - 2\hess_G(\sG)_*D_hG_*
\end{align}
\end{widetext}
At the minimum, Equation~\ref{hessG} becomes
\begin{align}
\hess_G\left(\sF_{t}\right)_* = \text{Hess}_G(\sG)L\text{Hess}_G(\sG)-\text{Hess}_G(\sG)
\end{align}
Finally, we note that in the presence of $h$, the minimum condition shifts to
\begin{align}
L(C)+G +h = 0
\end{align}
Differentiating this with respect to $h$ allows us to see
\begin{align}
D_hG_* = \left(L\hess_G(\sG)_* - \bbI\right)^{-1}
\end{align}
which gives, in conjunction with Equation~\ref{Hhess}
\begin{align}
\chi_0 = -\frac{1}{M}\Tr\left[\hess_G(\sG)_*\left(L\hess_G(\sG)_* - \bbI\right)^{-1}\right]
\end{align}
This expression readily exhibits the promised singularity at $T = \lambda_m$ for each individual disorder instance.
\bibliography{ref}

\begin{thebibliography}{18}%
\makeatletter
\providecommand \@ifxundefined [1]{%
 \@ifx{#1\undefined}
}%
\providecommand \@ifnum [1]{%
 \ifnum #1\expandafter \@firstoftwo
 \else \expandafter \@secondoftwo
 \fi
}%
\providecommand \@ifx [1]{%
 \ifx #1\expandafter \@firstoftwo
 \else \expandafter \@secondoftwo
 \fi
}%
\providecommand \natexlab [1]{#1}%
\providecommand \enquote  [1]{``#1''}%
\providecommand \bibnamefont  [1]{#1}%
\providecommand \bibfnamefont [1]{#1}%
\providecommand \citenamefont [1]{#1}%
\providecommand \href@noop [0]{\@secondoftwo}%
\providecommand \href [0]{\begingroup \@sanitize@url \@href}%
\providecommand \@href[1]{\@@startlink{#1}\@@href}%
\providecommand \@@href[1]{\endgroup#1\@@endlink}%
\providecommand \@sanitize@url [0]{\catcode `\\12\catcode `\$12\catcode
  `\&12\catcode `\#12\catcode `\^12\catcode `\_12\catcode `\%12\relax}%
\providecommand \@@startlink[1]{}%
\providecommand \@@endlink[0]{}%
\providecommand \url  [0]{\begingroup\@sanitize@url \@url }%
\providecommand \@url [1]{\endgroup\@href {#1}{\urlprefix }}%
\providecommand \urlprefix  [0]{URL }%
\providecommand \Eprint [0]{\href }%
\providecommand \doibase [0]{http://dx.doi.org/}%
\providecommand \selectlanguage [0]{\@gobble}%
\providecommand \bibinfo  [0]{\@secondoftwo}%
\providecommand \bibfield  [0]{\@secondoftwo}%
\providecommand \translation [1]{[#1]}%
\providecommand \BibitemOpen [0]{}%
\providecommand \bibitemStop [0]{}%
\providecommand \bibitemNoStop [0]{.\EOS\space}%
\providecommand \EOS [0]{\spacefactor3000\relax}%
\providecommand \BibitemShut  [1]{\csname bibitem#1\endcsname}%
\let\auto@bib@innerbib\@empty
\bibitem [{\citenamefont {Sachdev}\ and\ \citenamefont {Ye}(1993)}]{SY}%
  \BibitemOpen
  \bibfield  {author} {\bibinfo {author} {\bibfnamefont {S.}~\bibnamefont
  {Sachdev}}\ and\ \bibinfo {author} {\bibfnamefont {J.}~\bibnamefont {Ye}},\
  }\href {\doibase 10.1103/PhysRevLett.70.3339} {\bibfield  {journal} {\bibinfo
   {journal} {Phys. Rev. Lett.}\ }\textbf {\bibinfo {volume} {70}},\ \bibinfo
  {pages} {3339} (\bibinfo {year} {1993})}\BibitemShut {NoStop}%
\bibitem [{\citenamefont {Kitaev}(2015)}]{kit_talk}%
  \BibitemOpen
  \bibfield  {author} {\bibinfo {author} {\bibfnamefont {A.}~\bibnamefont
  {Kitaev}},\ }\href@noop {} {\bibfield  {journal} {\bibinfo  {journal} {Talks
  given at the KITP program: Entanglement in Strongly Correlated Quantum
  Matter, April 7 and May 27}\ } (\bibinfo {year} {2015})}\BibitemShut
  {NoStop}%
\bibitem [{\citenamefont {Sachdev}(2015)}]{subir1}%
  \BibitemOpen
  \bibfield  {author} {\bibinfo {author} {\bibfnamefont {S.}~\bibnamefont
  {Sachdev}},\ }\href {\doibase 10.1103/PhysRevX.5.041025} {\bibfield
  {journal} {\bibinfo  {journal} {Phys. Rev. X}\ }\textbf {\bibinfo {volume}
  {5}},\ \bibinfo {pages} {041025} (\bibinfo {year} {2015})}\BibitemShut
  {NoStop}%
\bibitem [{\citenamefont {Maldacena}\ and\ \citenamefont
  {Stanford}(2016)}]{MS}%
  \BibitemOpen
  \bibfield  {author} {\bibinfo {author} {\bibfnamefont {J.}~\bibnamefont
  {Maldacena}}\ and\ \bibinfo {author} {\bibfnamefont {D.}~\bibnamefont
  {Stanford}},\ }\href {\doibase 10.1103/PhysRevD.94.106002} {\bibfield
  {journal} {\bibinfo  {journal} {Phys. Rev. D}\ }\textbf {\bibinfo {volume}
  {94}},\ \bibinfo {pages} {106002} (\bibinfo {year} {2016})}\BibitemShut
  {NoStop}%
\bibitem [{\citenamefont {Klebanov}\ and\ \citenamefont
  {Tarnopolsky}(2017)}]{PhysRevD.95.046004}%
  \BibitemOpen
  \bibfield  {author} {\bibinfo {author} {\bibfnamefont {I.~R.}\ \bibnamefont
  {Klebanov}}\ and\ \bibinfo {author} {\bibfnamefont {G.}~\bibnamefont
  {Tarnopolsky}},\ }\href {\doibase 10.1103/PhysRevD.95.046004} {\bibfield
  {journal} {\bibinfo  {journal} {Phys. Rev. D}\ }\textbf {\bibinfo {volume}
  {95}},\ \bibinfo {pages} {046004} (\bibinfo {year} {2017})}\BibitemShut
  {NoStop}%
\bibitem [{\citenamefont {Witten}(2016)}]{witten2016syklike}%
  \BibitemOpen
  \bibfield  {author} {\bibinfo {author} {\bibfnamefont {E.}~\bibnamefont
  {Witten}},\ }\href@noop {} {\enquote {\bibinfo {title} {An syk-like model
  without disorder},}\ } (\bibinfo {year} {2016}),\ \Eprint
  {http://arxiv.org/abs/1610.09758} {arXiv:1610.09758 [hep-th]} \BibitemShut
  {NoStop}%
\bibitem [{\citenamefont {Kitaev}\ and\ \citenamefont
  {Suh}(2018)}]{kitaev2018soft}%
  \BibitemOpen
  \bibfield  {author} {\bibinfo {author} {\bibfnamefont {A.}~\bibnamefont
  {Kitaev}}\ and\ \bibinfo {author} {\bibfnamefont {S.~J.}\ \bibnamefont
  {Suh}},\ }\href@noop {} {\bibfield  {journal} {\bibinfo  {journal} {Journal
  of High Energy Physics}\ }\textbf {\bibinfo {volume} {2018}},\ \bibinfo
  {pages} {183} (\bibinfo {year} {2018})}\BibitemShut {NoStop}%
\bibitem [{\citenamefont {Polchinski}\ and\ \citenamefont
  {Rosenhaus}(2016)}]{polchinski}%
  \BibitemOpen
  \bibfield  {author} {\bibinfo {author} {\bibfnamefont {J.}~\bibnamefont
  {Polchinski}}\ and\ \bibinfo {author} {\bibfnamefont {V.}~\bibnamefont
  {Rosenhaus}},\ }\href {\doibase 10.1007/JHEP04(2016)001} {\bibfield
  {journal} {\bibinfo  {journal} {Journal of High Energy Physics}\ }\textbf
  {\bibinfo {volume} {2016}} (\bibinfo {year} {2016}),\
  10.1007/JHEP04(2016)001}\BibitemShut {NoStop}%
\bibitem [{\citenamefont {Bi}\ \emph {et~al.}(2017)\citenamefont {Bi},
  \citenamefont {Jian}, \citenamefont {You}, \citenamefont {Pawlak},\ and\
  \citenamefont {Xu}}]{bi_xu}%
  \BibitemOpen
  \bibfield  {author} {\bibinfo {author} {\bibfnamefont {Z.}~\bibnamefont
  {Bi}}, \bibinfo {author} {\bibfnamefont {C.-M.}\ \bibnamefont {Jian}},
  \bibinfo {author} {\bibfnamefont {Y.-Z.}\ \bibnamefont {You}}, \bibinfo
  {author} {\bibfnamefont {K.~A.}\ \bibnamefont {Pawlak}}, \ and\ \bibinfo
  {author} {\bibfnamefont {C.}~\bibnamefont {Xu}},\ }\href {\doibase
  10.1103/PhysRevB.95.205105} {\bibfield  {journal} {\bibinfo  {journal} {Phys.
  Rev. B}\ }\textbf {\bibinfo {volume} {95}},\ \bibinfo {pages} {205105}
  (\bibinfo {year} {2017})}\BibitemShut {NoStop}%
\bibitem [{\citenamefont {Georges}\ \emph {et~al.}(2001)\citenamefont
  {Georges}, \citenamefont {Parcollet},\ and\ \citenamefont {Sachdev}}]{GPS}%
  \BibitemOpen
  \bibfield  {author} {\bibinfo {author} {\bibfnamefont {A.}~\bibnamefont
  {Georges}}, \bibinfo {author} {\bibfnamefont {O.}~\bibnamefont {Parcollet}},
  \ and\ \bibinfo {author} {\bibfnamefont {S.}~\bibnamefont {Sachdev}},\ }\href
  {\doibase 10.1103/PhysRevB.63.134406} {\bibfield  {journal} {\bibinfo
  {journal} {Phys. Rev. B}\ }\textbf {\bibinfo {volume} {63}},\ \bibinfo
  {pages} {134406} (\bibinfo {year} {2001})}\BibitemShut {NoStop}%
\bibitem [{\citenamefont {Gross}\ and\ \citenamefont
  {Rosenhaus}(2017)}]{gross2017generalization}%
  \BibitemOpen
  \bibfield  {author} {\bibinfo {author} {\bibfnamefont {D.~J.}\ \bibnamefont
  {Gross}}\ and\ \bibinfo {author} {\bibfnamefont {V.}~\bibnamefont
  {Rosenhaus}},\ }\href@noop {} {\bibfield  {journal} {\bibinfo  {journal}
  {Journal of High Energy Physics}\ }\textbf {\bibinfo {volume} {2017}},\
  \bibinfo {pages} {93} (\bibinfo {year} {2017})}\BibitemShut {NoStop}%
\bibitem [{\citenamefont {Maldacena}\ \emph {et~al.}(2016)\citenamefont
  {Maldacena}, \citenamefont {Stanford},\ and\ \citenamefont {Yang}}]{MS2}%
  \BibitemOpen
  \bibfield  {author} {\bibinfo {author} {\bibfnamefont {J.}~\bibnamefont
  {Maldacena}}, \bibinfo {author} {\bibfnamefont {D.}~\bibnamefont {Stanford}},
  \ and\ \bibinfo {author} {\bibfnamefont {Z.}~\bibnamefont {Yang}},\ }\href
  {\doibase 10.1093/ptep/ptw124} {\bibfield  {journal} {\bibinfo  {journal}
  {Progress of Theoretical and Experimental Physics}\ }\textbf {\bibinfo
  {volume} {2016}},\ \bibinfo {pages} {12C104} (\bibinfo {year}
  {2016})}\BibitemShut {NoStop}%
\bibitem [{\citenamefont {Sachdev}(2010)}]{subir2}%
  \BibitemOpen
  \bibfield  {author} {\bibinfo {author} {\bibfnamefont {S.}~\bibnamefont
  {Sachdev}},\ }\href {\doibase 10.1103/PhysRevLett.105.151602} {\bibfield
  {journal} {\bibinfo  {journal} {Phys. Rev. Lett.}\ }\textbf {\bibinfo
  {volume} {105}},\ \bibinfo {pages} {151602} (\bibinfo {year}
  {2010})}\BibitemShut {NoStop}%
\bibitem [{\citenamefont {Jensen}(2016)}]{jensen}%
  \BibitemOpen
  \bibfield  {author} {\bibinfo {author} {\bibfnamefont {K.}~\bibnamefont
  {Jensen}},\ }\href {\doibase 10.1103/PhysRevLett.117.111601} {\bibfield
  {journal} {\bibinfo  {journal} {Phys. Rev. Lett.}\ }\textbf {\bibinfo
  {volume} {117}},\ \bibinfo {pages} {111601} (\bibinfo {year}
  {2016})}\BibitemShut {NoStop}%
\bibitem [{\citenamefont {Edwards}\ and\ \citenamefont
  {Anderson}(1975)}]{EASG}%
  \BibitemOpen
  \bibfield  {author} {\bibinfo {author} {\bibfnamefont {S.~F.}\ \bibnamefont
  {Edwards}}\ and\ \bibinfo {author} {\bibfnamefont {P.~W.}\ \bibnamefont
  {Anderson}},\ }\href {http://stacks.iop.org/0305-4608/5/i=5/a=017} {\bibfield
   {journal} {\bibinfo  {journal} {Journal of Physics F: Metal Physics}\
  }\textbf {\bibinfo {volume} {5}},\ \bibinfo {pages} {965} (\bibinfo {year}
  {1975})}\BibitemShut {NoStop}%
\bibitem [{\citenamefont {Marinari}\ \emph {et~al.}(2000)\citenamefont
  {Marinari}, \citenamefont {Parisi}, \citenamefont {Ricci-Tersenghi},
  \citenamefont {Ruiz-Lorenzo},\ and\ \citenamefont {Zuliani}}]{RSBREV}%
  \BibitemOpen
  \bibfield  {author} {\bibinfo {author} {\bibfnamefont {E.}~\bibnamefont
  {Marinari}}, \bibinfo {author} {\bibfnamefont {G.}~\bibnamefont {Parisi}},
  \bibinfo {author} {\bibfnamefont {F.}~\bibnamefont {Ricci-Tersenghi}},
  \bibinfo {author} {\bibfnamefont {J.~J.}\ \bibnamefont {Ruiz-Lorenzo}}, \
  and\ \bibinfo {author} {\bibfnamefont {F.}~\bibnamefont {Zuliani}},\ }\href
  {\doibase 10.1023/A:1018607809852} {\bibfield  {journal} {\bibinfo  {journal}
  {Journal of Statistical Physics}\ }\textbf {\bibinfo {volume} {98}} (\bibinfo
  {year} {2000}),\ 10.1023/A:1018607809852}\BibitemShut {NoStop}%
\bibitem [{\citenamefont {Tao}(2012)}]{taotopics}%
  \BibitemOpen
  \bibfield  {author} {\bibinfo {author} {\bibfnamefont {T.}~\bibnamefont
  {Tao}},\ }\href {https://books.google.com/books?id=Hjq\_JHLNPT0C} {\emph
  {\bibinfo {title} {Topics in Random Matrix Theory}}},\ Graduate studies in
  mathematics\ (\bibinfo  {publisher} {American Mathematical Soc.},\ \bibinfo
  {year} {2012})\BibitemShut {NoStop}%
\bibitem [{\citenamefont {White}(1980)}]{het-fit}%
  \BibitemOpen
  \bibfield  {author} {\bibinfo {author} {\bibfnamefont {H.}~\bibnamefont
  {White}},\ }\href {\doibase 10.2307/1912934} {\bibfield  {journal} {\bibinfo
  {journal} {Econometrica}\ }\textbf {\bibinfo {volume} {48}} (\bibinfo {year}
  {1980}),\ 10.2307/1912934}\BibitemShut {NoStop}%
\end{thebibliography}%

\end{document}